\documentclass[numberedappendix]{emulateapj}
\usepackage{color}
\usepackage{natbib}
\usepackage{epsfig}
\usepackage{amsmath}
\usepackage{subfigure}
\usepackage{booktabs} 
\newcommand{\ra}[1]{\renewcommand{\arraystretch}{#1}}
\def\icarus{\ref@jnl{Icarus}}          

\slugcomment{Accepted for publication In The Astrophysical Journal}

\shorttitle {Type I Planet migration in a Magnetized Disk. II.}
\shortauthors{Bans, K\"onigl, \&  Uribe}

\begin{document}

\title{Type I Planet Migration in a Magnetized Disk.\\ II. Effect of Vertical Angular Momentum Transport}
 \author{Alissa Bans,\altaffilmark{1} Arieh K\"onigl, and Ana Uribe}
\affil{Department of Astronomy and Astrophysics, University of Chicago, Chicago IL 60637, USA; abans@uchicago.edu}
\altaffiltext{1}{present address: Astronomy Department, Adler Planetarium, Chicago, IL 60605, USA}

\begin{abstract}
We study the effects of a large-scale, ordered magnetic field in
protoplanetary disks on Type I planet migration using a linear
perturbation analysis in the ideal-MHD limit. We focus on wind-driving
disks, in which a magnetic torque $\propto B_{0z} \partial
B_{0\varphi}/\partial z$ (where $B_{0z}$ and $B_{0\varphi}$ are the
equilibrium vertical and azimuthal field components) induces vertical
angular momentum transport. We derive the governing differential
equation for the disk response and identify its resonances and turning
points. For a disk containing a slightly subthermal, pure-$B_{0z}$
field, the total 3D torque is close to its value in the 2D limit but
remains lower than the hydrodynamic torque. In contrast with the 2D
pure-$B_{0\varphi}$ field model considered by \citet{Terquem03}, inward
migration is not reduced in this case when the field amplitude decreases
with radius. The presence of a subdominant $B_{0\varphi}$ component
whose amplitude increases from zero at $z=0$ has little effect on the
torque when acting alone, but in conjunction with a $B_{0z}$ component
it gives rise to a strong torque that speeds up the inward migration by
a factor $\gtrsim 200$. This factor could, however, be reduced in a real
disk by dissipation and magnetic diffusivity effects. Unlike all
previously studied disk migration models, in the $B_{0z}+\partial
B_{0\varphi}/\partial z$ case the dominant contributions to the torque
add with the same sign from the two sides of the planet. We attribute
this behavior to a new mode of interaction wherein a planet moves inward
by plugging into the disk's underlying angular momentum transport
mechanism.

\end{abstract}


\section{Introduction}
\label{intro}

Type I migration is the radial drift induced in a planet that is embedded in a protoplanetary disk by its linear gravitational interaction with the surrounding gas. This process, which is applicable to low-mass planets (typically up to Neptune's mass), is of great relevance to theories of planet formation and evolution (see, e.g., \citealt{LubowIda10}, \citealt{KleyNelson12}, \citealt{BaruteauMasset13}, and \citealt{BaruteauEtal13} for recent reviews). Early treatments of this interaction focused on isothermal disks with a smooth density distribution and inferred rapid inward migration. The subsequent discovery of ``hot Jupiters,'' giant planets orbiting within $\sim 10$ stellar radii of their host stars, has exacerbated the conundrum elicited by this finding. These  large, mostly gaseous, planets must have formed much farther out in their natal protoplanetary disks, where, according to the currently favored core-accretion model, they needed to assemble a large-enough ($\gtrsim$ a few times $M_\earth$) rocky core before a fast gas accretion phase (which created their envelopes) was triggered. However, the rapid inward drift predicted by the early models of Type I migration would prevent the postulated embryonic cores from lingering long enough at their formation sites for consistency  with the observed distribution of hot Jupiters \citep[e.g.,][]{IdaLin08}.

In the companion paper (\citealt{UribeEtal14}, hereafter Paper~I) we briefly review some of the more recent attempts to study Type I migration in the context of more realistic protoplanetary disk models, and we note a few of the promising results that have already been obtained. These include the effects of a disk magnetic field, which could be either a small-scale, disordered field associated with magnetohydrodynamic (MHD) turbulence (in particular, the turbulence induced by the magnetorotational instability) or a large-scale, ordered field that permeates the disk. These two field configurations likely play a central role in, respectively, the radial and vertical angular momentum transport in protoplanetary disks \citep[e.g.,][]{Balbus11,KoniglSalmeron11}. The effects of a small-scale, disordered field on planet migration include the stochastic motions induced in the smallest planets by MHD turbulence, the effective viscosity that prevents saturation of the corotation torque arising from the co-orbital region (which can be the dominant component of the torque on a planet undergoing Type I migration in a nonisothermal disk), 
and the magnitude of the mass threshold for opening a gap in the disk  (i.e., for the onset of Type II migration), which depends on the value of the effective viscosity. A novel feature of the presence of a large-scale, ordered field is the appearance of new types of waves that propagate in the disk (specifically, slow- and fast-magnetosonic [SMS and FMS, respectively], as well as Alfv\'en), which supplant the sound waves that characterize a hydrodynamic (HD) disk. The MHD waves modify the way in which the disk responds to the gravitational perturbation induced by the planet, and, in turn, the way in which the disk acts back on the planet to cause it to migrate. The studies conducted so far have demonstrated that a purely azimuthal field in 2D can slow down, and even reverse, inward migration if the magnetic field pressure is not much smaller than the thermal pressure and if the field amplitude decreases with radius fast enough \citep{Terquem03,FromangEtal05}. It was also shown that, while a strong azimuthal field suppresses the corotation torque, a weaker one modifies it in a way that can potentially lead to outward migration \citep[e.g.,][]{GuiletEtal13}. In the case of a purely vertical field it was found by \citet{MutoEtal08} that the only effect in 2D is the replacement of  the HD sound waves by the MHD FMS waves, with a resulting reduction in the magnitude of the torque, but that, in 3D, SMS and Alfv\'en waves also play a role (although these authors could not determine the net torque since their calculations were carried out in the shearing-sheet approximation). 

As we observed in Paper~I, a field configuration that is either purely azimuthal (and uniform with height) or purely vertical, as assumed in the aforementioned papers, is not representative of real protoplanetary disks. Such disks are likely permeated by open field lines that correspond to the interstellar field that originally threaded the parent molecular cloud cores and was dragged inward when the cores underwent gravitational collapse and formed the central stars and their associated circumstellar disks. The radially advected field lines assume a pinched (or hourglass) morphology and, in addition, are twisted by the differential rotation in the disk. In general, the field would thus have vertical, radial, and azimuthal components, which, near the disk surfaces, could have comparable magnitudes. This morphology is conducive to the formation of centrifugally driven outflows that can efficiently transport angular momentum away from the disk. This picture is supported by observations of protostellar disks as well as by theoretical considerations (see, e.g., \citealt{KoniglSalmeron11} and references therein, as well as \citealt{BansKonigl12}). Despite the strong azimuthal shear in the (typically Keplerian) protoplanetary disk, the magnetic field morphology in the giant-planet formation region can be expected to maintain an equilibrium structure on a timescale that greatly exceeds the local rotation period because the bulk of the gas at that distance (a few~AU) is generally weakly ionized and therefore magnetically diffusive. In the simplest representation of such an equilibrium open-field configuration, the field has an even symmetry about the midplane ($z=0$ in cylindrical coordinates $\{r,\varphi,z\}$), with the equilibrium (subscript `0') radial ($B_{0r}$) and azimuthal ($B_{0\varphi}$) field components changing sign there (but with the vertical component $B_{0z}$ remaining nonzero). 

The aim of Paper~I and this paper is to conduct a systematic investigation of the effects of a multi-component, ordered, large-scale magnetic field on Type~I planet migration. The ultimate goal of this study is to model a planet that is embedded in a wind-driving disk, which, as the discussion above indicates, involves all three spatial components of the field and must, for self-consistency, be treated within the framework of nonideal MHD. Given the complexity of this problem, its full treatment is deferred to a future publication, and the present study concentrates on simpler field configurations that can be investigated using ideal MHD. We employ the complementary approaches of a linear perturbation analysis (the focus of this paper) and of numerical simulations (the main focus of Paper~I). The linearization procedure is formulated in full generality in Section~\ref{methods} of this paper, the numerical procedure employed in calculating the net torque on the planet is described in Section~\ref{Nscheme}, and semianalytic solutions in~2 and~3 dimensions for three representative field configurations are presented in Section~\ref{results}. The most general field configuration that can be treated self-consistently within the framework of ideal MHD involves a field that has both vertical and azimuthal components that can vary with $r$ but not with $z$. This case is discussed extensively in Paper~I (which, in Section~I.2, also presents analytic results obtained from our perturbation analysis for this case).\footnote{We henceforth preface section, equation, and figure numbers in Paper~I by the numeral `I'.}  The 3 examples considered in this paper are a pure-$B_z$ field, a pure-$B_\varphi$ field whose amplitude increases with height (from zero at $z=0$), and the two combined. The first two configurations are treated numerically in Paper~I. The third configuration, which induces vertical angular momentum transport, mimics the situation in wind-driving 
 disks. It is, however, not entirely self-consistent in the context of ideal MHD since the radial velocity that results from the loss of angular momentum would give rise to a radial magnetic field component that, in the absence of magnetic diffusivity, would be rapidly sheared out of equilibrium by the differential rotation in the disk. This example is, nevertheless, included here in the hope that it could provide useful insights into the more general problem.\footnote{A similar motivation underlies the discussion of a vertical gradient of $B_r$ in Section~\ref{BrGrad}.} Finally, Sections~\ref{discussion} 
 (complemented by an appendix) and~\ref{conclusion} provide a discussion and a recapitulation, respectively, of the results presented in this paper.

\section{Linear Analysis} \label{methods}

\subsection{Basic Equations} \label{basic}

The dynamics of a magnetized protoplanetary disk is governed by the momentum conservation, mass conservation, and induction equations, given respectively by

\begin{equation} \rho \,  \Big [ \frac{\partial \textbf{v} }  { \partial t}  + ( \textbf{v} \cdot \nabla)  \textbf{v}  \Big ] = - \nabla  P + \textbf{F} -  \rho \nabla \psi \; , \label{eom} \end{equation}

\begin{equation}   \frac{\partial \rho}{\partial t} + \nabla \cdot (\rho \textbf{v })  =0\; ,  \label{con} \end{equation}
and

\begin{equation}  \frac {\partial \textbf{B} } {\partial t} + \nabla \times (\textbf{v} \times \textbf{B} )  =0\; , \label {ind} \end{equation}
where $\rho$ is the mass density, $\psi=GM_{\rm p}/\vert r - r_{\rm p} \vert$ the gravitational potential (with $M_{\rm p}$ and $r_{\rm p}$ being the planet's mass and orbital radius, respectively, and $G$ the gravitational constant), $\textbf{v}$ the velocity field, \textbf{B} the magnetic field, and $\textbf{F}$ the Lorentz force density

\begin{equation} \textbf{F}= (\nabla \times \textbf{B}) \times \textbf{B}/\mu  \label{loren} \end{equation} 
(with $\mu\equiv 4\pi$). The field is required to satisfy the solenoidal condition $\nabla \cdot \textbf{B} = 0$.

The induction equation was written in its ideal-MHD form, which deserves a clarification. As already noted in Section~\ref{intro}, the giant-planet formation zone in protoplanetary disks is expected to be weakly ionized and hence magnetically diffusive. This observation applies in particular to the disk midplane, where the planets reside, since this region is most strongly shielded from the main ionization sources of the disk (external cosmic rays, X-rays, and UV radiation). As a result, the ideal-MHD limit can be employed only under certain constraints. Since the focus of the current study is the effect of the magnetic field on the torque that the disk exerts on the planet (which arises from the nonaxisymmetric perturbations that the planet induces in the disk), it seems reasonable to require that the effective coupling time between the neutral disk component (which dominates the gravitational interaction with the planet) and the local magnetic field be shorter than the Keplerian rotation period (since any nonaxisymmetric perturbations would be washed out over timescales much larger than the inverse of the Keplerian angular velocity $\Omega_{\rm K}$). The ratio of these two timescales is quantified by the {\it Elsasser number}\/ $\Lambda \equiv v_{\rm A}^2/\Omega_{\rm K}\eta_\perp$, where $v_{\rm A}^2 = B^2/\mu\rho$ is the square of the Alfv\'en speed and $\eta_\perp$ is the magnetic diffusivity perpendicular to the field \citep[e.g.,][]{KoniglEtal10}. It can be expected that a magnetized disk would influence planet migration in an ideal-MHD--like way if $\Lambda \gg 1$ at the midplane.  Perhaps not surprisingly, this is also the minimum coupling condition for sustaining angular momentum transport through either MRI-induced turbulence or a centrifugally driven wind \citep[e.g.,][]{KoniglSalmeron11}. It is conceivable that the diffusivity of a real disk could give rise to interesting effects that cannot be captured with the current formulation,  but it nevertheless seems worthwhile to study the simpler ideal-MHD limit first.

Throughout this work we use an inertial, nonrotating, cylindrical coordinate system $(r,\varphi, z)$ centered on the star. The adopted field geometry and equilibrium disk conditions are described below. We also adopt an isothermal equation of state, $P = c^{2} \rho $, where $c$ is the isothermal speed of sound. Although our focus is on the disk midplane, we allow for vertical wave propagation, which entails taking account of the 3D structure of the disk \citep[e.g.,][]{TanakaEtal02,MutoEtal08}. We simplify the treatment by averaging over the effective thickness $2h$ of the disk (assumed to be $\ll r$ at the location of the planet) while imposing an even field symmetry. In this approach, the density $\rho$ is replaced by the surface density $\Sigma$, although we will also be referring to the vertically averaged equilibrium density $\rho_{\rm av} = \Sigma_0/2h$.

Following \citet{MutoEtal08}, we further simplify our treatment by neglecting the effects of the vertical component of the stellar gravitational field on the disk structure.

In the remainder of this paper we normalize the spatial variables $r$, $z$, and $h$ by the orbital radius of the planet $r_{\rm p}$, frequencies by the planet's orbital frequency $\Omega_{\rm p}$, time by $\Omega_{\rm p}^{-1}$, and velocities by $r_{\rm p} \Omega_{\rm p}$.

\subsection{Equilibrium Conditions}\label{Equilibrium}

We adopt a simplified equilibrium magnetic field configuration intended to capture the basic geometry of an even-symmetry open magnetic field that threads a geometrically thin disk:

\begin{equation} \textbf{B}_0 =\{ B_{0r}({z}), B_{0\varphi}({z}), B_{0z} ({r})  \}\; .    \end{equation}      
The geometrical thinness $h\ll 1$ implies, by the solenoidal condition, that $B_{0z}$ inside the disk is very nearly constant with height. Under the assumed even symmetry, the radial and azimuthal field components vanish at $z=0$ and increase in magnitude on going away from the midplane (with opposite signs in the two hemispheres). Based on a simplified treatment of a wind-driving disk \citep[e.g.,][]{WardleKonigl93}, we approximate $B_{0\varphi}({z}) =  B_{\varphi 0}\, {z}$, where $B_{\varphi 0}$ is a constant, and similarly for $B_{0r}({z})$. Since the dominant contribution to $B_\varphi$ in such disks is the shearing of $B_r$ by the differentially rotating gas, $B_{\varphi 0}$ generally has the opposite sign of $B_{r 0}$. We henceforth concentrate on the case where $B_{0z}$ and $B_{0r}$ are $> 0$ and $B_{0\varphi}$
is $<0$.\footnote{\label{note1}Note that our ansatz implies $B_{\varphi 0}=\partial B_{\varphi}/\partial {z}$, so the constant $B_{\varphi 0}$ can be used interchangeably with $\partial B_{\varphi}/\partial {z}$ when referring to the effect of a vertical gradient of the azimuthal field component. We distinguish this effect from that of an azimuthal field that is finite at the midplane and independent of ${z}$, which is considered in Paper~I.} We adopt a power-law form $B_{0z}({r})=B_{z0}{r}^{q_{z}}$ (with $B_{z0}$ a constant) for the radial dependence of the vertical field component, which is motivated by self-similar models of wind-driving disks \citep[e.g.,][]{BP82,ContopoulosLovelace94}. The other equilibrium magnetic field components would in general also vary with ${r}$, but we neglect this dependence in this paper. (It is, however, incorporated into the functional form we adopt for $B_{0 \varphi}$ in Paper~I.)

Using the above field morphology in Equation \eqref{loren}, we obtain for the vertically averaged equilibrium Lorentz force
\begin{multline}  \mu \langle\textbf{F} \rangle =  \bigg (  B_{r0}  B_{z0}{r}^{q_{z}} - q_{z} B_{z0}^{2} {r}^{2q_{z}-1}  -  B_{\varphi 0}^{2}  \frac{ h^2 }{3{r}} \ ,  \\ B_{\varphi 0}  B_{z0}{r}^{q_{z}} + B_{\varphi 0} B_{r0} \frac{h^2 }{3{r}}     \ , \  0   \bigg ) \ , \label{F} \end{multline} 
where the triangular brackets $\langle\rangle$ denote a vertically averaged (between $-h$ and $h$) quantity. The first two terms in the radial component of this expression represent the magnetic tension and pressure forces, 
respectively, which act to reduce the inward pull of gravity. (The third radial term is generally much smaller due to the $h^{2}$ factor.)
As we noted in Section~\ref{intro}, the strong shearing of a radial field component cannot be self-consistently incorporated into an equilibrium ideal-MHD model, so, in the applications given in this paper and in Paper~I, we set $B_{0r}=0$. This eliminates the radial tension, which would typically dominate the radial Lorentz force in a real wind-driving disk. The first (and dominant) term in the azimuthal component of Equation\eqref{F} represents the braking torque acting on the disk. When substituted into the $\varphi$ component of Equation \eqref{eom} (which describes the conservation of angular momentum), we obtain an expression for the induced accretion velocity $v_{0r}$,

\begin{equation} v_{0r}  =\frac{ 2\Omega \langle F_{\varphi} \rangle} {\rho_{av}\kappa^2}\ ,      \label{Vor} \end{equation} 
where $\Omega$ is the angular velocity and $\kappa$ is the epicyclic frequency ($\kappa^{2} = \frac{2\Omega}{r} \frac{\partial (r^{2}\Omega)}{\partial r}$). 

Equation \eqref{Vor} highlights the fact that a disk threaded by a vertical magnetic field and possessing a nonzero vertical gradient of the azimuthal magnetic field would experience vertical angular momentum transport. This is a key property of disks that drive centrifugal outflows. Because of the strong vertical gradients that characterize geometrically thin disks, this transport could be very efficient. In fact, when such a wind is launched, the resulting accretion speed (assuming adequate magnetic coupling, $\Lambda >> 1$, in the disk) is of the order of $c$ \citep[e.g.,][]{SalmeronEtal07}, which is much higher than the speeds that arise from radial angular momentum transport by a small-scale, disordered field. As was noted in Section~\ref{intro}, the field-line bending induced by the inflowing gas would give rise to a radial field component that, in turn, would lead to a rapidly growing azimuthal field component and a departure from equilibrium unless magnetic diffusivity were taken into account. Since we intend to include nonideal-MHD effects in a future work, we retain the $v_{0r}$ term in the basic formulation
presented in this paper and assume an equilibrium velocity field of the form $\textbf{v}= (v_{0r}, r \Omega, 0)$ in the ensuing perturbation analysis. However, the semianalytic results that we obtain are derived in the limit $v_{0r}
=0$, which should be an adequate approximation given that $c/r\Omega$ (which is of the order of $h$) is $\ll 1$.\footnote
{A possible exception to this conclusion is our treatment of the case $B_{0z} \partial B_{0 \varphi} /\partial {z} \ne 0$ in Section~\ref{BzBphiResults2}, which, as was already pointed out in Section~\ref{intro}, does not handle the equilibrium magnetic field configuration in a self-consistent manner.}

\subsection{The Disk Response}\label{PerturbSet}

To explore how our equilibrium disk model responds to the perturbing potential of the planet, $\psi'$ (normalized by $r_{\rm p}^2\Omega_{\rm p}^2$), we follow the method of \citet{Terquem03} and earlier analyses by considering small perturbations and linearizing the basic equations. We allow the planet to perturb the disk in all 3 directions, and Fourier transform in time as well as in space along the $\hat{\varphi}$ and $\hat{z}$ directions  by considering all Fourier quantities to be of the form $X^\prime_{m,k_z}(r) e^ { i (  m \varphi + k_{z} z - \omega t )}$, with the frequency taken to be $\omega = m\,\Omega_{\rm p}$. As in \citet{Terquem03}, we denote all perturbed quantities with a prime. The planet's potential itself can be expanded in a Fourier series, $\psi^\prime = \sum_{m=0}^\infty{\psi^\prime_m \cos{m\phi}}$, where $\phi\equiv\varphi-  \Omega_{\rm p} t $, and expressed in terms of the generalized Laplace coefficients $b^{m}_{1/2}(\alpha)$ \citep[e.g.,][]{Ward89,KP93,Terquem03}:

\begin{equation}
\psi^\prime_m = \left\{
\begin{array}{lcc}
-\frac{M_{\rm p} }{M_{*} r}\delta_m b^{m}_{1/2}(\alpha) \quad \ \ \ \ r>1\\
-\frac{M_{\rm p} }{M_{*}}\  \delta_m b^{m}_{1/2}(\alpha) \quad  \ \ \ \ r<1\; ,
\end{array} \right.
\label{poten}
\end{equation}
where $M_*$ is the stellar mass (assumed to be $\gg M_{\rm p}$), the coefficient $\delta_m$ is equal to 0.5 for $m=0$ and to 1.0 for all the other (positive integer) values of $m$,
\begin{equation}  b^{m}_{1/2}(r) = \frac{2}{\pi} \int_0^\pi  \frac{\cos{m\phi} \ {\rm d}\phi} { ( q^{2} + p^{2}
\alpha^{2} - 2\alpha \cos{\phi} )^{1/2}}\ , \label{LaplaceC}  \end{equation}
and where, for $r>1$, $\alpha = 1/r$, $q=1$, and $p^2=1+\epsilon^2$; whereas for $r<1$, $\alpha = r$, $p=1$, and $q^2 = 1 + \epsilon^2$ (with $\epsilon$, a real number of magnitude $\ll 1$, being the potential softening parameter; see Equation~(I.20)).

Linearizing equations \eqref{eom} and \eqref{con} yields
\begin{multline} i m \sigma v_{r}^\prime  - 2v_{\varphi}^\prime\Omega + \bigg( v_{0r} \frac{\partial v_{r}^\prime}{ \partial r}  +  v_{r}^\prime \frac {\partial v_{0r}}{\partial r}  + v_{z}^\prime \frac{\partial v_{or}}
{\partial z} \bigg ) = \\ - \frac {\partial}{\partial r} \big ( \psi^\prime + W^\prime) + \frac{2h}{\Sigma_0} \big( -\frac{W^\prime}{c^{2}} \langle F_{r} \rangle + \langle F_{r}' \rangle \big)  \label{eomr'}\; ,
\end{multline}
\begin{multline}  i m \sigma  v_{\varphi}^\prime + \frac{\kappa^{2}}{2 \Omega}v_{r}^\prime + v_{0r}\bigg(\frac{\partial v_{\varphi}^\prime}{\partial r} + \frac{v_{\varphi}^\prime}{r} \bigg ) =  -\frac{ im}{r} \big( \psi^\prime + W^\prime \big) \\ +\frac{2h}{\Sigma_0} \big( -\frac{W^\prime}{c^{2}}\langle F_{\varphi} \rangle + \langle F_{\varphi}^\prime \rangle \big)  \label{eomphi'} \; , \end{multline}
\begin{equation}  i m \sigma_0  v_{z}^\prime  + v_{0r}\frac{\partial v_{z}^\prime}{\partial r} = - i k_{z} W^\prime - i k_{z} \psi^\prime + 2h\frac{\langle F_{z}^\prime \rangle}{\Sigma_0} \label {eomz'} \; ,  \end{equation} 
and
\begin{multline} \frac{i m \sigma W^\prime} {c^{2}} =   \frac {-1}{r\Sigma_0} \frac{\partial}{\partial r} \big(r \Sigma_0 v_{r}^\prime \big ) \\ - \frac{im}{r} v_{\varphi}^\prime - i k_{z} v_{z}^\prime -  \frac {1}{r\Sigma_0} \frac{\partial}{\partial r} \big(r \Sigma^\prime v_{0r} \big )\; , \label{con'}\end{multline}
where $W^\prime={\Sigma^\prime c^{2}}/{\Sigma_0}$ is the enthalpy perturbation, $\sigma \equiv \Omega-\Omega_{p}$ ($=\Omega - 1$ in normalized units), and where we suppressed the Fourier labels ($m$ and $k_z$) on the primed quantities to simplify the presentation. To calculate the perturbed Lorentz force components in Equations \eqref{eomr'} -- \eqref{eomz'}, we use 
\begin{equation}  \mu \textbf{F}^\prime= (\nabla \times \textbf{B}^\prime ) \times \textbf{B}_0 + \big( \nabla \times \textbf{B}_0 \big ) \times \textbf{B}^\prime  \label{F'CF}\end{equation}
and the following expression from \citet{ChandraFermi53},
\begin{equation} \textbf{B}^\prime = \nabla \times ( \boldsymbol{\xi}  \times \textbf{B}_0 )\; , \label{B'CF}\end{equation}
which relates the perturbed magnetic field to the Lagrangian displacement $\boldsymbol {\xi}$. This relation expresses the ideal-MHD condition of the magnetic field lines being ``frozen'' into the matter, and can be derived from the induction equation in the limit of a negligible equilibrium velocity. Equation \eqref{B'CF} does not strictly hold when $v_{0r}$ is finite, but its use in the present analysis is justified by the fact that the results we derive are obtained in the limit $v_{0r}
= 0$.

Note that the vertical averaging in the perturbation equations is done \emph{after} all the spatial derivatives are evaluated. Thus, for example, the perturbed vertical momentum equation contains a term $\propto \partial B_{0\varphi}/\partial z$ (if it is nonzero) even though the equilibrium vertical magnetic force vanishes in our model (see Equation \eqref{F}). It is also worth noting that, in general, the perturbation equations do not contain terms involving the spatial derivative of the equilibrium density if the temperature is assumed to be a spatial constant (as is done in this paper, following, e.g., \citealt{KP93} and \citealt{Terquem03}). Consequently, the linearization results are not sensitive to our neglect of a vertical density gradient (induced by either tidal gravity or a vertical magnetic pressure gradient) in the equilibrium momentum equation.

The Lagrangian displacement is related to the Eulerian velocity perturbations through
\begin{equation} \textbf{v}^\prime= \frac{\partial \boldsymbol{\xi }}{\partial t} + (\textbf{v}_0 \cdot \nabla) \boldsymbol{\xi}  - (\xi \cdot \nabla) \textbf{v}_0      \label{delv'}\end{equation}
\citep[e.g.,][]{ZhangLovelace05}. Thus, 
continuing from now on to suppress the Fourier labels on the perturbed quantities,
\begin{equation}v_{r}^\prime= i m \sigma \xi_{r}   + v_{0r}  \frac{\partial \xi_{r} }{ \partial r}   - \frac{\partial v_{or} }{\partial r} \xi_{r}\ ,  \label{vr'} \end{equation}
\begin{equation}v_{\varphi}^\prime= i m \sigma \xi_{\varphi}  - r \xi_{r} \frac{\partial \Omega} { \partial r}   + v_{0r}  \frac{\partial \xi_{\varphi} }{ \partial r}   - \frac{ v_{0r} \xi_{\varphi} }{r}\ ,  \label{vphi'}\end{equation}
\begin{equation}v_{z}^\prime= i m \sigma \xi_{z}   + v_{0r}  \frac{\partial \xi_{z} }{ \partial r}\ .  \label{vz'} \end{equation} 
Substituting these expressions into Equations \eqref{eomr'} -- \eqref{con'} and carrying out the necessary algebra, one obtains a second-order differential equation for $\xi_{r}$ of the form
\begin{equation} A_{2} (r)\frac{\partial^{2} \xi_{r}}{\partial r ^{2}} + A_{1} (r) \frac{\partial \xi_{r}}{\partial r } + A_{0} (r) \xi_{r}  = S_{1} (r) \frac {\partial \psi^\prime}{\partial r} + S_{0} (r) \psi^\prime\ , \label{Diff} \end{equation}
which can be integrated numerically (see Section~\ref{Nscheme}).\footnote {The functions $A_0$, $A_1$, $A_2$, $S_0$, and $S_1$ that appear as coefficients in Equation \eqref{Diff} are provided as online supplementary material in the form of both a Mathematica notebook and a PDF document.}
The derived solution for $\xi_{r}$ can then be plugged back into the linearized equations of motion and continuity to obtain $W^\prime$ and to evaluate the torque exerted by the planet on the disk. 

\subsection{Resonances and Turning Points}\label{ResTP}

The solution of Equation \eqref{Diff} becomes singular at the locations where the leading-order coefficient, $A_{2}(r)$, goes to zero. These locations correspond to the resonances of the differential equation and are manifested by a divergence of the density perturbation. Their presence also signals the appearance of distinct regions of wave propagation in the disk. The torque exerted in the regions surrounding the resonances can potentially dominate the response of the disk and thus determine the rate and direction of planet migration. It is therefore important to calculate the resonant locations and examine their contribution to the integrated torque. 

In general, the differential equation \eqref{Diff} describes wave propagation in the disk, and we can rewrite a homogeneous version of it in terms of just the second- and zeroth-order terms. We do this, following \citet{Terquem03}, by defining $\lambda_{r}\equiv \xi_{r} \text{ exp} \bigg ( \frac {1}{2}  \int \frac{A_{1}}{A_{2}} \partial r \bigg )$ as a new dependent variable, which yields
\begin{equation}  \frac {\partial^{2} \lambda_{r}}{\partial r^{2}}  + K\lambda_{r}  = 0\; , \label{lam} \end{equation}
where 
\begin{equation} K= \frac{A_{0}}{A_{2}} - \frac{1}{4} \bigg( \frac{A_{1}}{A_{2}} \bigg)^{2}  - \frac{1}{2} \frac{\partial}{\partial r} \bigg(\frac{A_{1}}{A_{2}} \bigg)    \ . \label{Kroot}\end{equation} 
When $K$ is real, Equation \eqref{lam} has wave-like solutions if $K>0$ and evanescent solutions if $K<0$. More generally, when $K$ is complex (with real and imaginary parts given by $\Re\{K\}$ and $\Im\{K\}$, respectively),
 this equation always has wave-like solutions, but whether or not the waves propagate is determined by the magnitude and signs of $\Re\{K\}$ and $\Im\{K\}$ (see Appendix~\ref{appen2} for details). In our model, $K$ is complex for disks that contain a vertically variable equilibrium field component (nonzero $\partial B_{\varphi}/\partial {z}$ or $\partial B_r/\partial z$), and real for vertically constant field configurations.
The locations where the solution changes its character from wave propagation to evanescence (i.e., where $K = 0 $) are known as the {\it turning points}\/ of the differential equation. Since it is the density wave propagation from the turning points that is responsible for the exchange of angular momentum with the planet \citep[e.g.,][]{GoldreichTremaine79}, finding the locations of the turning points 
makes it possible to identify the regions of the disk that exert a back torque 
on the planet. In the next subsection we evaluate the locations of the resonances and turning points of a simplified version of 
Equation \eqref{Diff} and show how their properties depend on the magnetic field configuration. 

\subsection{ Two- and Three-Dimensional Modes} \label{ResTPresults}

With a nonzero $v_{0r}$, Equations \eqref{eomr'} -- \eqref{con'} all contain both zeroth-order and derivative terms of the perturbed variables, which means that it is impossible to eliminate all variables besides $\xi_{r}$ from the equations. As discussed in Section \ref{Equilibrium}, $v_{0r}$ is an effectively small velocity, so in this initial study we drop all the terms in which it appears in the equations. Equation \eqref{B'CF} then manifestly satisfies the linearized induction equation, and the 
(vertically averaged) 
perturbed field is given by
\begin{multline} \langle \textbf{B}' \rangle= \big ( -B_{r 0}  \xi _{z}+i  k_{z} B_{z0} {r}^{q_{z}} \xi_{r} , B_{\varphi 0} \xi_{z} + ik_{z}B_{z0}{r}^{q_{z}}\xi_{\varphi} , \\  - \frac{im B_{z0}{r}^{q_{z}}\xi_{\varphi}}{{r}} - \frac{(q_{z} + 1)  B_{z0}{r}^{q_{z} } \xi_{r}}{{r}}  - B_{z0}{r}^{q_{z}}\frac{\partial \xi_{r}}{\partial r} \;    \big ) \ . \label{B'} \end{multline}

It can be seen from Equation \eqref{F'CF} that $F_{\varphi}'$ includes a term of the form $ B_{r 0}  {\partial \xi_{
\varphi}}/{\partial r} $, which would result in the inclusion of a derivative of $\xi_\varphi$ in Equation \eqref{eomphi'} and thereby make the system \eqref{eomr'}--\eqref{con'} difficult to solve. The leading coefficient in this case does not have a simple closed form that can be used to extract the resonances. Therefore, in the interest of maximizing our insight into the relevant resonances, we further simplify the treatment by letting $B_{0r}$ 
be zero. Since the appearance of a radial field component is tied physically to the development of a vertically-sheared radial velocity field in the disk (in particular, when $B_z$ and $\partial v_{or}/\partial z$ are nonzero; see Equation \eqref{ind}), this approximation is consistent with our neglect of the $v_{0r}$ terms.

With these two simplifications, the real part of the leading coefficient $A_2(r)$ of the resulting second-order differential equation has two obvious roots, which yield two resonance conditions. The first one is 
\begin{equation}  \resizebox{0.98\hsize}{!}{$  m^{2}\sigma^{2}= \frac{k_{z}^{2} c^{2} v_{{\rm A}z}^{2} + \frac{h^{2}}{3} \bigg(  v_{{\rm A}z}^{2} \partial v_{{\rm A}\varphi}^{2} (k_{z}^{2} + \frac{m^{2}}{r^{2}} ) +  \partial v_{{\rm A}\varphi}^{2} c^{2} \frac{ m^{2} }{ r^{2} } \bigg)   }{c^{2} + v_{{\rm A}z}^{2} +  \frac{h^{2}}{3} \partial v_{{\rm A}\varphi}{2}  }  $} \label{MRNobrg}  \\\ . \end{equation}
Here $v_{{\rm A}z}$ is the (rescaled) Alfv\'{e}n velocity associated with the ${z}$ component of the field, $v_{{\rm A}z}^2=\frac{ B_{z0}^{2} }{\mu\Sigma} $, and $ \partial v_{{\rm A}\varphi}^{2}\equiv\frac{ (\partial B_{\varphi} / \partial z )^{2}}{\mu\Sigma} = \frac{ B_{\varphi 0}^{2}} {\mu\Sigma}$.\footnote{Note that $B_{\varphi 0}$ does \emph{not} represent the midplane value of $B_\varphi$, which is zero in our model.}
Ignoring order-$h^{2}$ terms, the above equation indicates that 
one type of resonances occurs at the locations (one interior and one exterior to the planet's radius) where the ``corotating'' perturbation frequency $m\sigma$ is equal to the frequency of a SMS wave propagating along $B_z$. As in Paper~I, we refer to these as ``Magnetic Resonances" (MRs).  The second resonance condition is
\begin{equation} m^{2}\sigma^{2}= k_{z}^{2}v_{{\rm A}z}^{2} - \partial v_{{\rm A}\varphi}^{2}\left( 1 -   \frac{h^{2}m^{2}}{3r^{2}}   \right)  \label{ARNobrg} \ . \end{equation}
If not for the vertical gradient in the $\hat{\varphi}$ component, Equation \eqref{ARNobrg} would indicate that there is another
type of resonance at the locations where the perturbation frequency in the corotating frame matches that of a shear Alfv\'{e}n wave propagating in the $\hat {z}$ direction. As in Paper~I, we refer to these as ``Alfv\'{e}n Resonances" (ARs).  Equations \eqref{MRNobrg} and \eqref{ARNobrg} indicate that the MRs and ARs for the field configuration considered in this paper differ from those derived for the multicomponent field explored in Paper~I (which is distinguished from the one discussed here by having a nonzero azimuthal component at the disk midplane). We find that the MRs depend strongly on the vertical field and very little on the azimuthal field gradient (which only appears in conjunction with factors of order $h^{2}$), whereas for the configuration considered In Paper~I the MR locations are sensitive also to the azimuthal field component (see Equation~(I.13)). Furthermore, in the limit of $k_{z} \ne 0$ and no vertical field, the ARs are defined by $m^{2}\sigma^{2}= -\partial v_{{\rm A}\varphi}^{2} (1 - \frac{h^2 m^2}{3 r^2} )$. Thus, the ARs in this case only exist for values of $m$ that satisfy $\frac{h^2 m^2}{3 r^2} > 1 $, in stark contrast to the case of an azimuthal field with $k_{z} \ne 0$ considered in Paper~I (see Equations~(I.11) and~(I.12)), in which the ARs always exist. However, the ARs disappear, irrespective of the field geometry, in the strictly 2D
limit (see Section~I.2.1). The 2D cases of $\partial B_{\varphi} / \partial z \ne 0$ do, however, exhibit weak MRs (see Equation \eqref{MRNobrg}), which is consistent with the results for a uniform azimuthal field \citep[e.g.,][]{Terquem03}, but they turn out to have a minimal effect on the net torque.

\begin{figure*}
\begin{center}
\includegraphics[ totalheight=.4\textheight, angle=0]{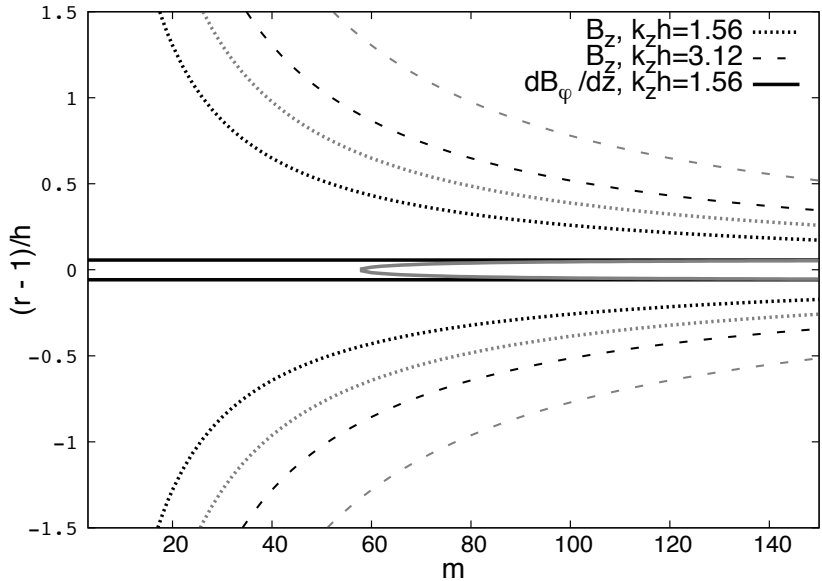}
\caption{ Locations of the magnetic (black lines) and Alfv\'{e}n (gray lines) resonances for two different field configurations, $\partial B_{\varphi}/\partial z =0, \; B_z\ne0$ and $\partial B_{\varphi}/\partial z\ne 0, \; B_z=0\;$ as a function of the azimuthal mode number $m$.   The  resonances for $B_{z}\ne 0$ are shown for both $k_{z}h=1.56$ and $k_{z}h=3.12$  (the solid and dashed lines respectively), but the $\partial B_{\varphi}/\partial z \ne 0$ case (dotted lines) is displayed at only one vertical wavenumber since the resonance locations for this latter case are independent of $k_{z}$ (see Equations \eqref{MRNobrg} and \eqref{ARNobrg}). For the pure-$B_{z}$ case, the resonances move away from the planet as the vertical wavenumber increases, implying that the contribution of these modes to the net torque decreases with $k_z$. The resonance locations for a field configuration with both $B_{z}\ne 0$ and  $\partial B_{\varphi}/ \partial z\ne 0$ are very similar to those for the pure-$B_{z}$ case, so this configuration is not shown separately. The  Alfv\'{e}n resonances for the $\partial B_{\varphi}/ \partial z\ne 0$  case only exist for $m\gtrsim55$.  \label{Reslocation}}
\end{center}
\end{figure*}

The dependence of the resonance locations on the field configuration and on the wavenumber $k_z$ is illustrated in Figure~\ref{Reslocation}. 
The MRs are shown by the black lines, and the ARs by the gray ones. For a pure vertical field, the ARs are always located outside the MRs (i.e., farther away from the corotation radius, where the disk's angular velocity is equal to that of the planet). Both resonances have locations that depend on the azimuthal wavenumber $m$, getting closer to the planet with increasing $m$. For a pure-$B_z$ field, both of these resonances move further away from the planet as $k_{z}$ increases, diminishing the contributions of the resonances to the torque. For the $B_{\varphi 0}\ne0,\; B_{z0}=0$ case and $k_{z}\ne 0$, the MRs are located symmetrically with respect to the corotation radius, and their radial position does not vary with $m$. This is similar to the behavior of the MR resonances for the 2D, pure-$B_\varphi$ case \citep[e.g.,][]{Terquem03}, except that in the case considered here the MRs only appear away from the midplane.\footnote{This can be seen in the numerical simulations presented in Figure~I.15. We capture the effect of these MRs even though our analysis is restricted to the midplane, where $B_\varphi$ is zero, because we use a vertically averaged value of $B_{\varphi 0}$.} In this case, an additional AR appears interior to the MR. As noted above, this AR only exists for wave numbers $m$ above a critical value that depends on the disk thickness. For a thin disk with $h \ll 1$, this critical value is large, which makes the torque contribution of these ARs negligible. 

\begin{figure*}
 \begin{center}
  \includegraphics[ totalheight=0.5\textheight, angle=0]{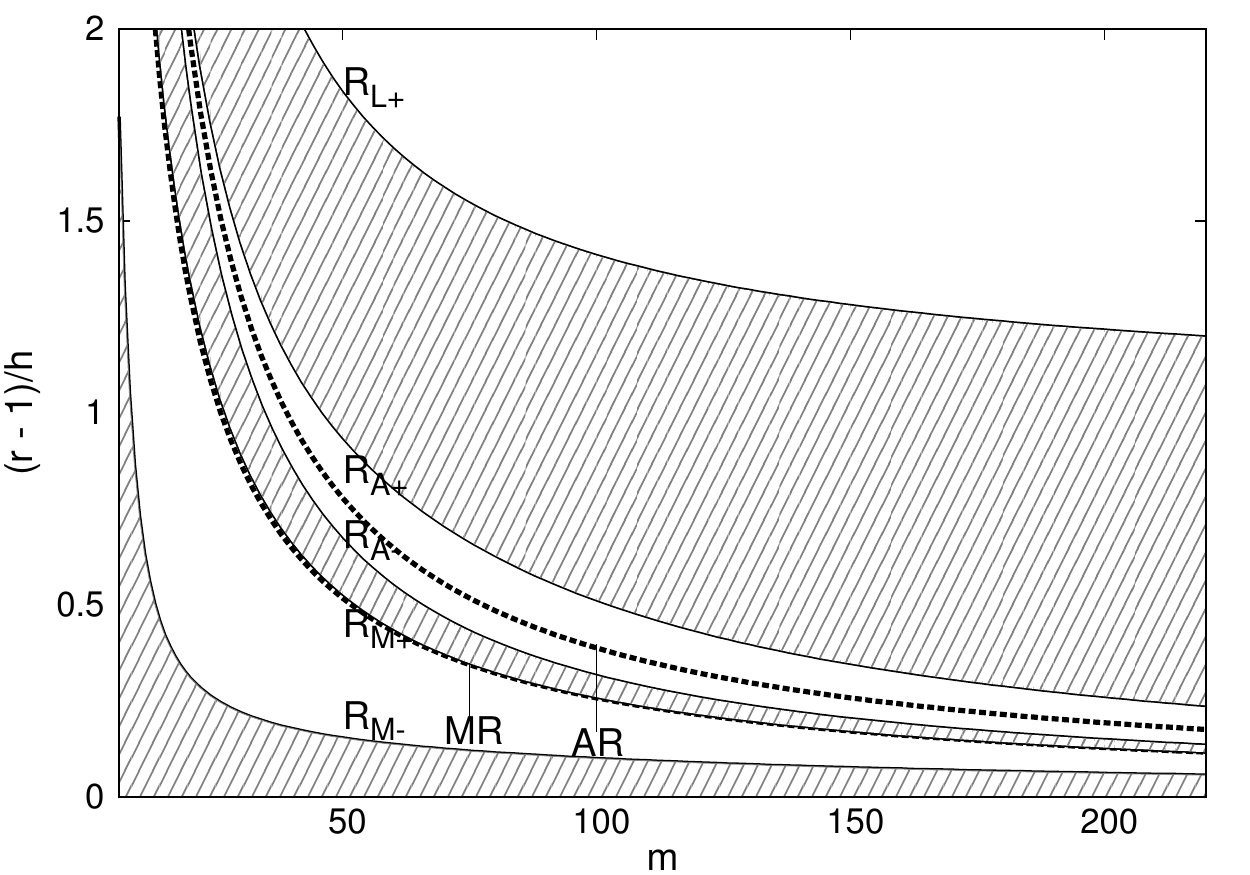}
\caption{Wave propagation (clear) and evanescence (shaded) regions outside the planet's orbit for a purely vertical field configuration with $\beta_{z} = 0.8$ as a function of the azimuthal mode number $m$. The vertical wavenumber is fixed at $k_{z}h = 1.56$. The magnetic resonance given by Equation \eqref{MRNobrg} is labeled 'MR', whereas the Alfv\'{e}n resonance given by Equation \eqref{ARNobrg} is labeled 'AR'. The turning points are labeled $R_{L+}$, $R_{A+}$, $R_{A-}$, $R_{ M+}$, and $R_{M-}$ (see text for details). \label{WavePropBz}}
  \end{center}
\end{figure*}

Figure~\ref{WavePropBz} shows the outer ($r>1$) wave propagation regions and their relation to the locations of the turning points and resonances for the case of a pure vertical field and a finite value of $k_{z}$. The magnetic field strength is parameterized by the value of $\beta_z \equiv c^2/v_{{\rm A}z}^2$, where $\beta$ is half the thermal-to-magnetic pressure ratio. (This ratio is generally $\gtrsim 1$ near the midplane of a magnetically well-coupled, wind-driving disk; e.g., \citealt{KoniglSalmeron11}.) Each of the resonances (MRs and ARs) is associated with a pair of turning points that straddle it. We label the turning points according to the resonances they surround. From the planet's location outward, there are four such turning points, ${\rm R_{M-}}$, ${\rm R_{M+}}$, ${\rm R_{A-}}$, and ${\rm R_{A+}}$, and then an additional one, ${\rm R_{L+}}$, the modified effective outer Lindblad resonance. Two regions of wave propagation, from which torque is exerted on the planet, surround each resonance and are bounded by the turning points. Waves also propagate away from the outer Lindblad turning point. The integrated torque exerted on the planet from the region $r>1$ depends on the properties of these three regions as well as on the point-like contributions from each of the resonances. The net torque, in turn, is determined by the balance between the opposing effects of the outer and inner disk regions, and by the contribution of the co-orbital region. Note that, as the value of the azimuthal mode number $m$ increases, the resonances move closer to the planet but their associated regions of wave propagation become narrower. It is thus not obvious a priori which mode numbers dominate the torque. We address this issue on the basis of our numerical results in Section~\ref{pureBz}.

\citet{MutoEtal08} studied the pure-$B_z$, $k_z \ne 0$ case using the shearing-sheet model and the WKB approximation. Similarly to what we find, they identified three regions of wave propagation, corresponding (in order of increasing distance from the planet) to the SMS, Alfv\'en, and FMS waves. Their Figures~1 and~2(c) (for which $\beta_z=0.9$) are evidently most closely related to the results we show in Figure~\ref{WavePropBz}. Their inner evanescence region, whose outer boundary (which they term LR-) corresponds to our ${\rm R_{M-}}$, 
exists only for $k_{z}^{2}v_{{\rm A}z}^{2}>3\,\Omega_{\rm p}^{2}$.\footnote{As was noted by \citet{MutoEtal08}, this condition is the same as the linear stability condition against the MRI \citep[e.g.,][]{Balbus11}.} In our work we also find that this region is present only when 
$k_{z}^2v_{{\rm A}z}^2$ is sufficiently large, although the lower bound that we obtain with our more general treatment ($3.04\,\Omega_{\rm p}^{2}$) is slightly different from the analytic limit. \citet{MutoEtal08} identified the inner boundary of the intermediate evanescence region with the locus of the magnetic resonances rather than with that of the turning points ${\rm R_{M+}}$ (which their derivation misses); we note in this connection that, even though we clearly distinguish between these two loci, in practice they are very close to each other in our example for all values of $m$. The inner boundary of the intermediate evanescence region, given by the turning points ${\rm R_{A-}}$, appears to correspond to what these authors call ``vertical resonances.''\footnote{Note that \citet{MutoEtal08} refer to genuine resonances as well as to turning points as ``resonances.'' It is thus worth reemphasizing that the locations where $K$ (Equation \eqref{Kroot}) vanishes, which include the classical effective Lindblad resonances in an unmagnetized disk, are actually turning points: They mark the onset of wave propagation in the disk, and, in contrast with actual resonances, they do not correspond to a divergence in the perturbed density.} Their derivation, however, also misses the turning points ${\rm R_{A+}}$, and thus they identify the inner boundary of the outer evanescence region with the locus of the Alfv\'en resonances. In our example, the Alfv\'en resonances are spatially well separated from the turning points ${\rm R_{A+}}$ except at low values of $m$, where the WKB approximation (which requires $m\ll k_z r$) applies. This comparison illustrates the limitations of the WKB approximation in providing an accurate description of the disk--planet interaction. The outer boundary of the outer evanescence region is, however, correctly identified by this approximation with the locus of the modified effective Lindblad resonances.

\begin{figure*}
\begin{center}
\includegraphics[ totalheight=.5\textheight, angle=270]{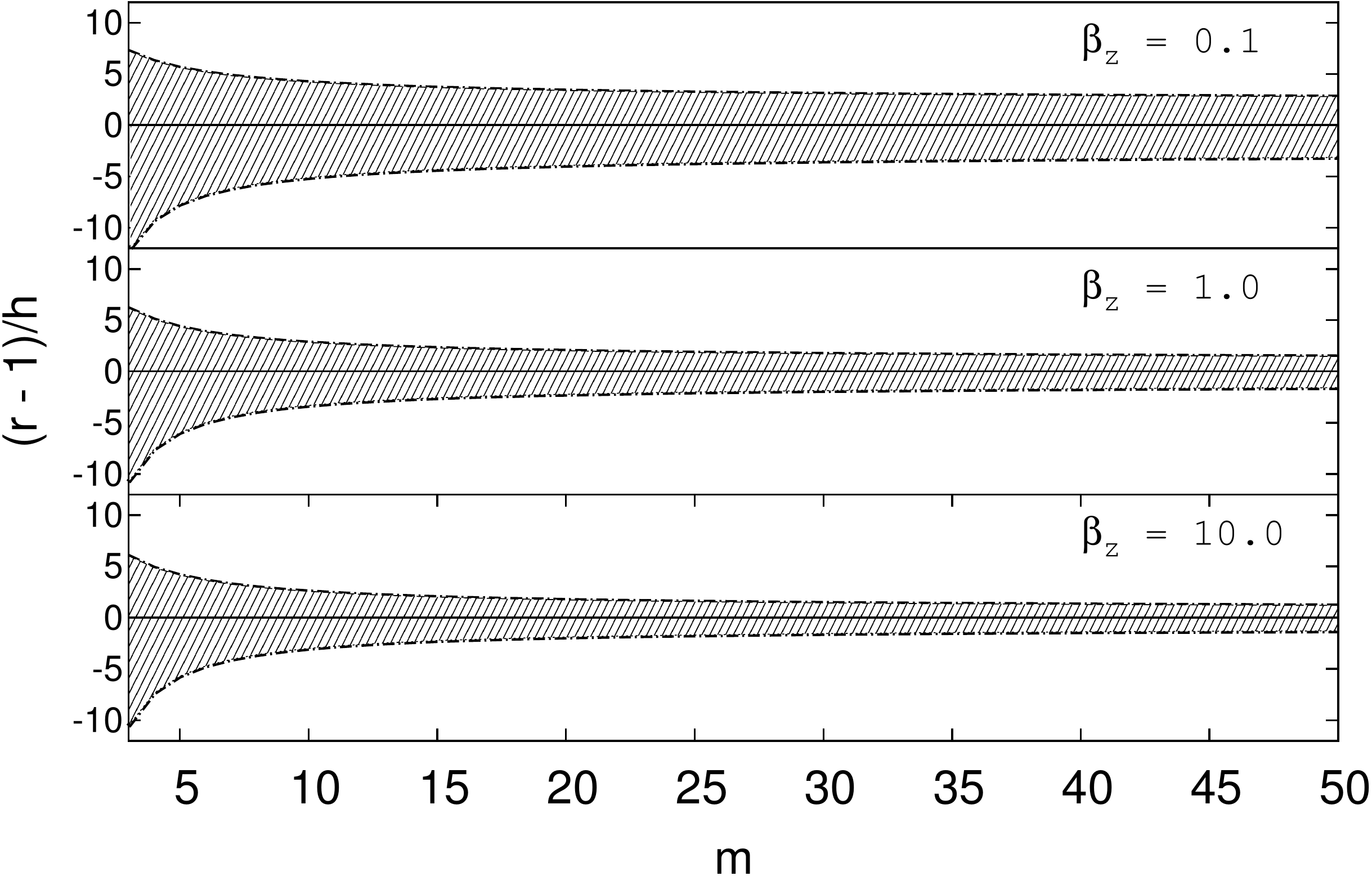}
\caption{Wave propagation (clear) and evanescence (shaded) regions on both sides of the planet's orbit for a purely vertical field in the 2D limit. The three panels correspond to different magnetic field strengths, parameterized by $\beta_z \equiv 
c^{2}/{v_{{\rm A}z}^{2}}$. Note the asymmetry between the inner and outer turning points (dashed lines) in each case.
\label{TPvbeta} }
\end{center}
\end{figure*}

In the limit $k_{z} =0$ of the pure-$B_z$ case, both resonances vanish and there is only one set of turning points --- the effective Lindblad resonances modified by the presence of the field, as described by Equation~(37) of \citet{MutoEtal08}. As noted above, our formalism does not involve the shearing-sheet approximation that was employed by these authors, and we are therefore able to evaluate the difference between the torques exerted by the inner and outer disk regions and hence the net torque acting on the planet. Figure~\ref{TPvbeta} shows the distances of the turning points on the two sides of the planet and their dependence on the field strength. It is seen that the basic structure is similar to that in a purely HD disk in that the outer effective Lindblad resonances, which exert a negative torque on the planet, lie closer to the planet than the inner turning points and can therefore be expected to dominate the interaction. This is true for all azimuthal mode numbers,  although the asymmetry is stronger for low values of $m$. Under these circumstances, the planet is predicted to migrate inward. As the field strength increases, the turning points move farther away from the planet, which suggests that the magnitude of the net torque, and hence the rate of inward migration, are reduced. This heuristic inference is borne out by our numerical results (see Figure~\ref{PBzTorque} below) as well as by the 2D numerical simulations presented in Paper~I (see Figure~I.3). We further verified by numerical simulations that the decrease in the net torque with increasing $B_z$ applies also in 3D.

When both $B_{\varphi 0}$ and $B_{z0}$ are nonzero,
the coefficients $A_{0}$, $A_{1}$, and $A_{2}$, and hence $K$ (Equation \eqref{Kroot}) are complex. In this case, we expect wave propagation when $\Re\{K\} > 0$ and $\Re\{K\} \gg | \Im\{K\} |$ as well as when $|\Im\{K\}| \gg |\Re\{K\}|$ (see Appendix~\ref{appen2}). Figure \ref{WavePropBphiG} shows the predicted regions of wave propagation (at a given $m$) for this field configuration for both 2D and 3D modes. The $k_{z} \ne 0$ modes behave similarly to the 3D pure-$B_{z}$ case, which can be understood from the fact that, even at the disk surface (subscript `s'), the amplitude of $B_\varphi$ remains a small fraction of $B_z$ ($|B_{\varphi \rm s}/B_{z0}|=0.13$ in this example). However, when $k_{z}=0$, there is an interior wave propagation region that starts very close to the planet (corresponding to a region of large  $|\Im\{K\}|$), which differs from the behavior of a pure-$B_{z}$ configuration in 2D. Although Figure~\ref{WavePropBphiG} only shows the wave propagation regions for a given value of $m$, we have verified that the area of the interior propagation region in both the 2D and 3D cases decreases with increasing $m$ (much as it does in the 3D pure-$B_z$ model shown in Figure~\ref{WavePropBz}). The presence of an inner wave-propagation region in 2D for this field configuration suggests that, in contrast with the pure-$B_z$ disk, the torque would be enhanced (rather than reduced) in this case in comparison with the HD limit (see Section~\ref{BzBphiResults2}).

\subsection{Effect of Vertical Gradient of  $B_{r}$}
\label{BrGrad}

As we already pointed out, a vertical gradient of the radial field component cannot be self-consistently incorporated into our ideal-MHD disk model. However, in view of the fact that the magnetic tension force ($\propto B_z\partial B_r/\partial z$) could play an important dynamical role in real wind-driving disks \citep[e.g.,][]{WardleKonigl93}, it is of interest to at least look for clues to its possible effect on planet migration. As we noted in Section~\ref{ResTPresults}, when $B_{r0}\ne 0$, one cannot obtain a differential equation in the form of  Equation \eqref{Diff}. We therefore attempt to apply an iterative approach to the procedure that yielded this equation. First, we ignore the $ B_{r0}{\partial \xi_{\varphi}}/{\partial r}$ term in Equation \eqref{eomphi'} and solve for  $\xi_{\varphi}$ as a function of  $\xi_{r}$ and its derivatives only.
We then differentiate the resulting expression for $\xi_{\varphi}$ and use it to evaluate the term we originally ignored so as to obtain an updated value for $\xi_{\varphi}$. We solve the rest of the system of equations using this updated value. In general, the leading-order coefficient of Equation \eqref{Diff} that is obtained in this way is rather complicated, and there is no simple form for its roots. However, one can obtain approximate resonance conditions by keeping the dominant terms in the limit $B_{\varphi 0} \rightarrow 0$ and also dropping the smaller-order $\propto h$ terms. In this way we find the following approximate conditions:
\begin{equation} m^{2}\sigma ^{2}  -  v_{{\rm A}z}^{2} k_{z}^{2}   - 2 v_{{\rm A}z} \partial v_{{\rm A}r}\frac{\partial \Omega / \partial r}{\sigma} =0\; ,  \label{brgr1}\end{equation}
\begin{equation} \begin{split} m^{2} \sigma^{2} ( v_{{\rm A}z}^{2} + c^{2} ) -  v_{{\rm A}z}^{2} c^{2} k_{z}^{2} + \partial v_{{\rm A}r}^{2}   ( v_{{\rm A}z}^{2} + c^{2} ) - \\  \partial v_{{\rm A}r} \frac{ \partial v_{{\rm A}z}}{\partial r} ( v_{{\rm A}z}^{2} + c^{2} ) \label{brgr2} =0\; , \end{split} \end{equation}
\noindent
where $\partial v_{{\rm A}r}$ is defined analogously to $\partial v_{{\rm A}\varphi}$ in Section~\ref{ResTPresults}. 

In the absence of a radial field, Equation \eqref{brgr1} reduces to the Alfv\'{e}n resonance condition and Equation \eqref{brgr2} to the magnetic resonance condition for a pure-$B_z$ configuration. However, Equation \eqref{brgr1} indicates that, when $\partial v_{{\rm A}r}\ne 0$, there is an inherent asymmetry between the inner and outer Alfv\'{e}n-like resonances. This equation is formally cubic in $\sigma$, and since $\partial \Omega / \partial r$ is negative, it yields an additional resonance at $r>1$.  This extra resonance persists even when $k_{z}=0$. 
This suggests that the presence of a vertical gradient of $B_{r}$ could enhance the torque exerted by waves exterior to the planet's orbit, which induces inward migration.  Although we cannot check on the validity of this prediction in the present study, we hope to pursue this case further in a future investigation.

\begin{figure}
\begin{center}
\includegraphics[ totalheight=0.5\textheight, angle=0]{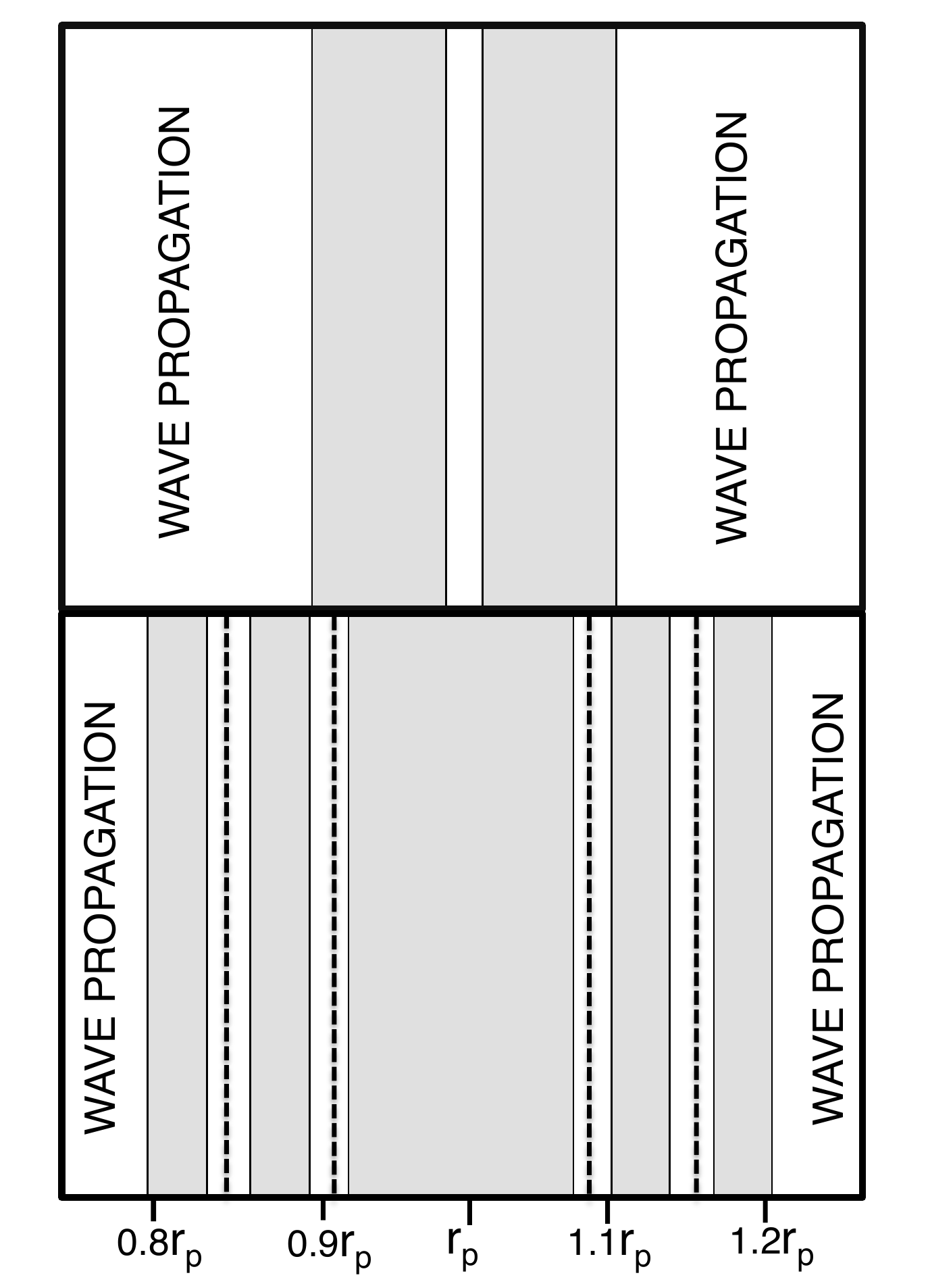}
\caption{Distribution of wave propagation (clear) and evanescence (shaded) regions in a disk with a $B_{z}\ne 0$, $\partial B_{\varphi}/ \partial z \ne 0$ field configuration 
for $k_zh=0$ (top) and $k_zh \ne 0$ (bottom). The 2D case corresponds to the same parameters as the left panel of Figure~\ref{BpGwave}, although in this schematic we neglect the contribution of the order-$h^2$ terms. The bottom panel corresponds to  the $k_{z}h=3.12$, $m=20$ case shown in Table~\ref{tab1}. The dashed lines indicate the locations of the Alfv\'{e}n and magnetic resonances. 
\label{WavePropBphiG}}
\end{center}
\end{figure}

\section{Numerical Scheme}\label{Nscheme}

To solve Equation \eqref{Diff}, we employ the same methodology used by \citet{KP93} and \citet{Terquem03}. To save computational time, we employ known recurrence formulas to evaluate the Laplace coefficients (Equation \eqref{LaplaceC}) for a given $m$. The relations we use are outlined in the appendix of \cite{KP93} and also in \cite{book1} and \cite{book2}. 

The integration of Equation \eqref{Diff} is generally tricky due to the steepness of the planet potential and the presence of the various resonances. To achieve the required accuracy while using a standard initial-value integration routine, it is necessary to adopt small steps and require high precision; here we apply a fifth-order Runge-Kutta method with adaptive stepsize control \citep{Press1992} and use 16-bit precision. To deal with any poles at $\sigma=0$, we follow \citet{KP93} and \citet{Terquem03} and displace the pole from the real axis by $ i\delta$, where $\delta$ is a real number with $|\delta| \ll 1$, i.e.,  $\sigma \rightarrow \sigma - i \delta $. This effectively also displaces the resonances from the real axis. We start the integration from the planet and integrate toward the inner and outer domain boundaries (specified below). We find that this methodology produces a smoother run than a ``monotonic''  integration from the inner boundary to the outer one.

At the boundaries of our integration domain,  where $r^{2}\sigma^{2} \gg c^{2}$,  we neglect $A_{1}$ in comparison to $A_{0}$, which is justified by the fact that, to leading order, $A_{0}/A_{1} \rightarrow (r^2\sigma^{2}/c^{2}) (m^{2} - \kappa^{2}/ \sigma^{2} )$.
 In this limit, the homogeneous version of Equation \eqref{Diff} becomes 
\begin{equation} A_{2} (r)\frac{\partial^{2} \xi_{r}}{\partial r ^{2}} + {A_{0}} (r) \xi_{r}  = 0   \label{Diffwkb}  \; , \end{equation} 
where we retain only the leading-order term in each coefficient. Applying the 
WKB approximation by assuming that the solution of Equation \eqref{Diffwkb} takes the form  $\xi_{r} \sim A e^{i k_{\rm W} r}$ yields $k_{\rm W}= {A_{0}}/A_{2}$ . For the case of a field with nonzero $B_{z}$ and $\partial B_{\varphi} / \partial z$,  $ {A_{0}}$ is given by
\begin{equation}   {A_{0}}= m^{4}\sigma^{4} \left( m^{2}\sigma^{2} - \kappa^{2} + \frac{ k_{z}^{2} c^{2} \kappa^{2}}{m^{2}\sigma^{2}} -k_{z}^{2}c^{2}  \right) \label{A0l} \; .\end{equation}
Although the terms that include $k_{z}$ are small in comparison with the other terms on the right-hand side of Equation \eqref{A0l}, they can influence the results at small values of $m$. Note also that, in the limit $B \rightarrow 0$ and $k_{z} \rightarrow 0$, ${A_{0}}/A_{2}$ reduces to the dispersion relation for acoustic density waves in a 2D HD disk. 

We use this WKB result as a boundary condition, $ \partial \xi_{r} / \partial r = i k_{\rm W} \xi{r}$, at the inner and outer edges of the integration domain. We still need initial values for $\xi_{r}$ and $ \partial \xi_{r} / \partial r$ to start the integration, so, similarly to the standard shooting method and following \citet{KP93} and \citet{Terquem03}, we start with random values and use an iterative approach to capture the true solution. For the sake of completeness, we provide a summary of this procedure below; additional details can be found in \citet{KP93}. 

Starting with random values implies that the solution obtained by the numerical scheme will in general contain a linear combination of solutions of the homogeneous equation (subscript `hs') and of the particular solution of the inhomogeneous equation (subscript `ps'), i.e., $ \xi_r= \xi_{\rm ps} + C_{1} \xi_{\rm hs1} + C_{2} \xi_{\rm hs2}$. To attain convergence toward the true solution, we simultaneously integrate the two linearly independent  homogeneous solutions, $\xi_{\rm hs1}$ and $\xi_{\rm hs2}$. After one full integration, we can use the boundary conditions to constrain the constants $C_{1}$ and $C_{2}$ (specifically, our integrated solutions for  $\xi_{\rm ps}, \xi_{\rm hs1} , \xi_{\rm hs2}$ and their derivatives yield a 2x2 system for $C_{1}$ and $C_{2}$ when plugged into $  \partial \xi_{r} / \partial r = i k_{\rm W} \xi_{r} $ at the disk boundaries). After solving for these constants, we perform one more integration, this time initializing $\xi_{\rm ps}$ as $\xi_{ps,{\rm old}} + C_{1} \xi_{\rm hs1} + C_{2} \xi_{\rm hs2} $ --- 
this allows $\xi_{\rm ps}$ to converge toward $\xi_{\rm r}$. As in \citet{Terquem03}, we find that good convergence is attained already after two iterations.

In contrast with the behavior of the 2D, purely azimuthal configuration considered by \citet{Terquem03}, in which, for example, the location of the magnetic resonance was independent of $m$, for the field configurations that we consider the locations of the resonances vary with the azimuthal mode number $m$ as well as with the vertical wavenumber $k_{z}$. In our case, large vertical wavenumbers and small azimuthal mode numbers correspond to resonances (and, more generally, wave propagation regions) that lie far from the planet. Calculating the torque can then be computationally expensive, as it requires integration over a large region to capture the full effect of the resonances. To overcome this challenge, we limit ourselves to low values of $k_z h$  (although we also calculate the torque for a few higher values of this product) and interpolate between some of the larger values of $m$. For very high values of $m$, we find it necessary to reduce the size of the integration domain (see Section~\ref{BzBphiResults2}). As was also the case in \citet{KP93}, we find that for certain disk parameters we cannot calculate the contribution of some low azimuthal mode numbers (the differential equation becomes stiff  and the needed step size to integrate is excessively small). In these cases we extrapolate to estimate the torque contribution from these low $m$. This, in turn, introduces an error into the summed (over $m$) torque value, but we find that the qualitative behavior of the solution is not affected by this procedure. 
Note that, as in \citet{KP93} and \citet{Terquem03}, we only attempt to calculate modes with $m\ge2$. For $m=0$, there is no effect on the torque since $T_{m}\propto m$ (see Equation \eqref{T_mA}), whereas for $m=1$, the results involve a contribution from the disk's outer edge \citep[see][]{Shuetal90}, which we eliminate by enforcing an outgoing wave condition at the radial boundaries.

Solutions to Equation \eqref{Diff} are highly oscillatory, leading to strong cancellations of the torque contributions from adjacent zones, particularly in the outer regions of the disk. For this reason, the numerical results can depend on the size of the integration domain. 
In our treatment, we use $r_{\rm in}=0.3\, r_{\rm p}$ and $r_{\rm out}=1.7\, r_{\rm p}$ as the integration boundaries for $m\ge6$.
This domain is slightly larger than the one used by \cite{KP93} and \cite{Terquem03} since we need to ensure that the integration region is sufficiently large also for the $k_{z} h\ne 0$ cases, in which resonances and turning points lie farther away from the planet. For $m< 6$ we adopt an even larger domain, $r_{\rm in}=0.2\, r_{\rm p}$ and $r_{\rm out}=5.0\, r_{\rm p}$.  We assume that the torque value for a given $m$ has effectively converged if the value does not differ by more than $\sim 1\%$ from the torque obtained using a slightly larger range. Following the practice in \citet{KP93} and \citet{Terquem03}, we set the values of the parameters $\delta$ and $\epsilon$ to be $1\times10^{-6}$ and $1\times10^{-4}$, respectively; however, our results are not sensitive to the specific choices of these values provided that they remain small (either one can be a factor of $\sim 10$ larger before any changes are detected in the results).

After solving for $\xi_{r}$, we obtain $W^\prime$ from Equations \eqref{eomr'} -- \eqref{con'} and proceed to calculate the torque (at a given $m$) per unit area ($A$) that is exerted by the planet using
\begin{equation} \frac{\partial T_{m}}{\partial A} = \frac{m}{2} 
\Re \left\{ \frac{ i W^\prime \Sigma_0 \psi^\prime}{c^{2}} \right\}\label{T_mA} \end{equation}
and 
\begin{equation} T_{m} = 2 \pi \int_{r_{in}}^{r_{out}}{  \frac{\partial T_{m}}{\partial A} r \partial r } \label{T_m}\end{equation}    
\citep[see][]{KP93}. To get the cumulative torque, we sum over $m$ (which in practice we do through an integral, assuming that a given value of $T_{m}$ characterizes a small interval of $m$ values). To obtain the total torque in the 3D cases, we approximate the additional sum over the vertical wave numbers by adding up the contributions of the specific $k_z$ values that we evaluate.

In this paper we use the convention, commonly adopted in linear analyses of the type we carry out \citep[e.g.,][]{KP93,Terquem03}, wherein one calculates the torque exerted {\it by}\/ the planet {\it on}\/ the disk. In this case,  a positive torque implies inward migration whereas a negative torque indicates outward migration. This sign convention is the {\it opposite}\/ of that commonly used in numerical studies, which concentrate on the effect that the disk has on the planet. The latter convention is also adopted in Paper~I, where the focus is on numerical simulations. We caution the reader to be mindful of this difference when comparing the results of the two papers.

\section{Results} \label{results}

\subsection{ Hydrodynamic Limit }
\label{hydro}

We ran our scheme with ${B} =0$ to check against the HD results of \citet{KP93}. In this limit, waves propagate away from the effective Lindblad resonances and there is a genuine resonance at the corotation radius. The total torque exerted by the planet is positive, representing the standard inward-migration result. We found that our results reproduce those of \citet{KP93} well. In particular, we obtained a good match to their plot (in Figure~2) of the real and imaginary parts of $W^\prime(r)$ for $m=10$, and our calculated cumulative dimensionless torque of $\sim 1300$ is in
good agreement with the value reported in their Figure~14. Our results similarly match the HD calculations presented in \citet{Terquem03}, and we further validated our numerical scheme by reproducing the azimuthal field results given in that paper (in particular, the plot of $T_m$ vs. $m$ for the uniform-field case, shown in the top panel of Figure~6 of that paper).\footnote{The 2D, pure-$B_\varphi$ case was calculated using the formulation  presented in Section~I.2.}

We also considered the 3D torque for this case. \citet{TanakaEtal02} originally found that the 3D torque is weaker than the 2D torque by a factor of $\gtrsim 2$. For the $k_{z}h$ values used in this paper, we found the ratio to be closer to $\sim 1$. We note, however, that an exact correspondence is not expected since \citet{TanakaEtal02} used different disk parameters and included vertical stratification. Furthermore, the value of the 3D torque in our analysis also depends on the range of $k_z h$ that we use (which in practice corresponds to the choice of $h$, the effective disk half-thickness). Our chosen minimum value of $k_{z} h$ ($=1.56$) is based on applying our adopted value of $\beta_z$ ($=0.8$) 
 to the condition for the existence of the inner evanescence region (or, equivalently, the condition for the existence of the innermost turning point) in the pure-$B_{z}$ case (see Section~\ref{ResTPresults}). Regarding the upper limit of the adopted range, it is in practice limited by the computational difficulties encountered for large values of $k_{z} h$ (See Section~\ref{Nscheme}). We were, however, able to explore the 3D torque in this case for sufficiently large values of this product (represented by ever thinner disks) to confirm that it is indeed weaker than the 2D HD torque, in qualitative agreement with the \citet{TanakaEtal02} result.

\subsection{$B_{z}\ne0$, $\partial B_\varphi/\partial z = 0$ Field Configuration}
\label{pureBz}

Although \citet{MutoEtal08} investigated the case of a pure-$B_z$ field, they formulated the problem in the shearing-sheet approximation and considered the 3D regime only for disks that are strongly magnetically dominated ($\beta_z\le10^{-2}$) and thus do not correspond to wind-driving systems. We therefore take a fresh look at this case.

\begin{figure*}
\begin{center}
\includegraphics[totalheight=.56\textheight,angle=0]{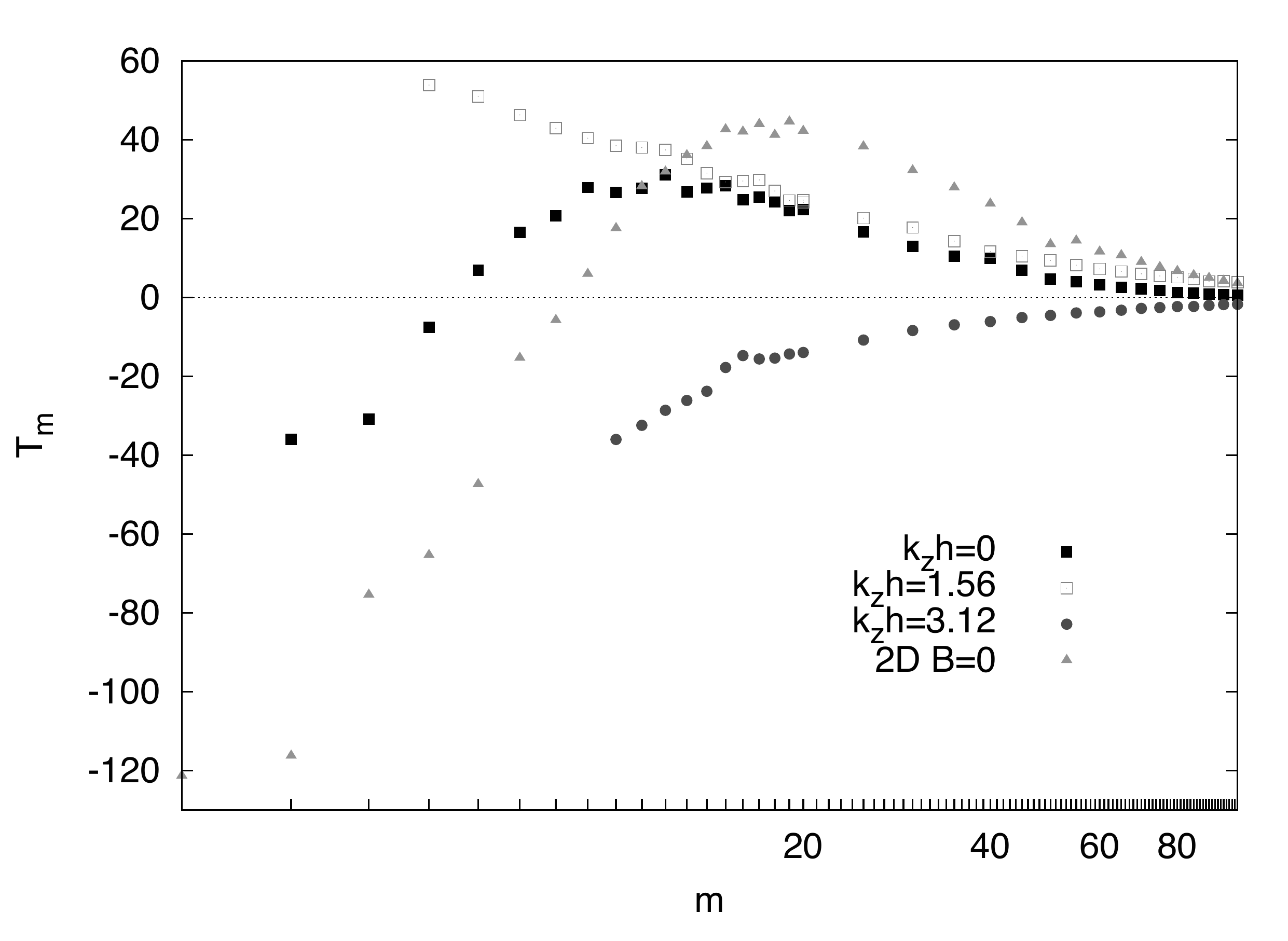}
\caption{Spatially integrated torque exerted by the planet on a thin ($h = c/(r_{\rm p}\Omega_{\rm p})=0.03$), uniform disk that is threaded by a purely vertical field (of strength $\beta_{z}=0.8$) as a function of the azimuthal mode number $m$. Results for the 2D limit ($k_{z} = 0$) and for two nonzero vertical wavenumbers are shown, with the 2D HD case included for reference. Because of the stiffness of the governing differential equation at low values of $m$, we were only able to obtain numerical results for $m>8$ for $k_{z}h=3.12$,  $m>4$ for $k_{z}h=1.56$, and $m>2$ for the $k_z h=0$ case. Extrapolating where necessary, the cumulative dimensionless torques for $k_{z}h=0$, 1.56, and 3.12 are $\sim 700$, 1400, and $-1100$, respectively. Neglecting the contribution from higher $k_{z}$ modes, we thus estimate the total dimensionless torque to be $\sim 1000$, 
a fraction $\sim 0.7$ of the 2D HD value ($\sim 1300$).}
\label{PBzTorque}
\end{center}
\end{figure*}

Figure~\ref{PBzTorque} shows the integrated torque $T_m$ as a function of $m$ for a uniform disk in the 2D limit and for two finite values of $k_z$. We also plot the HD result for comparison. The 2D integrated torque is invariably lower than in the unmagnetized case, resulting in a cumulative dimensionless torque for the chosen disk parameters that is a factor 
$\sim 2$ smaller than the HD value. As was already pointed out by \citet{MutoEtal08} (see also Section~\ref{ResTPresults} and Paper~I), this reduction can be attributed to the outward shift of the effective Lindblad resonances in a magnetized disk, which increases with the field strength (see Figure~\ref{TPvbeta}). The behavior of the 3D modes is more complicated. For $k_zh = 1.56$, $T_m$ is positive and generally larger than the corresponding value for $k_z = 0$. This is due to the appearance in this case of the magnetic and Alfv\'en resonances and of their associated regions of wave propagation. However, for the larger vertical wavenumber considered in this figure ($k_zh=3.12$), the net torque for each value of $m$ is {\it negative}. This is a consequence of the fact --- revealed by a close examination of Figure~\ref{Reslocation} --- that the inner MR lies slightly closer to the planet than the outer MR. This asymmetry becomes more pronounced, with a corresponding increase in its effect, as $k_z$ increases. We thus expect the integrated torque to remain negative as $k_zh$ grows further; however, its contribution to the total torque would be smaller than that of the $k_zh=3.12$ mode because the resonances and associated wave-propagation regions move outward with increasing $k_z$. We estimate the total torque in this case by adding up the net torques for the three values of $k_z$ shown in Figure~\ref{PBzTorque}; we find a value
 that is still smaller ($\sim 0.7$) than the 2D HD case shown in the same plot. This ballpark estimate is in good agreement with the total torque obtained in the numerical simulations presented in Paper~I (see Figure~I.10). In Section 3.4.1 of that work we point out that, for the adopted model parameters, our semianalytic estimates yield comparable values for the 3D and the 2D HD torques. Hence, our conclusion that the torque in a disk with a pure-$B_{z}$ field is roughly half that of an unmagnetized disk applies in both 2D and 3D. 

The breakdown of the contributions to the integrated torque from different regions in the disk for $k_zh=3.12$,  $ \beta_z=0.8 $, and two values of $m$ (20 and 100) is shown in Table~\ref{tab1}. For $m=20$, we also show results for a field that is twice as strong ($\beta_z=0.4$). The regions are chosen as follows (see Figure~\ref{WavePropBz} for the meaning of the different labels): from the inner/outer edge of the disk to the inner/outer $R_{L+}$ ($T_{Li}$ and $T_{Lo}$, respectively); from the inner/outer $R_{L+}$ to the inner/outer $R_{A-}$ ($T_{ARi}$ and $T_{ARo}$, respectively), and from the inner/outer $R_{A-}$ to the planet's location ($T_{MRi}$ and $T_{MRo}$, respectively). 
These sectors encompass the propagation regions of FMS, Alfv\'en, and SMS waves, respectively. Our ability to distinguish between the contributions of the regions on the inner and outer sides of the planet helps us gain insights that could not be obtained with a shearing-sheet model.

The comparison between the $m=20$ and $m=100$ results for $\beta_z=0.8$ reveals that the contribution of the magnetic resonance region dominates, and its magnitude increases with $m$, on each of the two (inner and outer) sides of the planet. However,  the contributions from the two sides are closer in magnitude, resulting in a smaller net torque, for the higher-$m$ case. On the other hand, the contribution from the 
Alfv\'en resonance region (from either side of the planet as well as the net one) decreases with $m$. These two trends combine to make the integrated torque much higher (with comparable contributions from the magnetic and Alfv\'en resonance regions) for $m=20$. This behavior also explains why $T_m$ peaks at a comparatively low value of $m$ ( see Figure~\ref{PBzTorque}). 

The torque arising from the regions around a resonance is a combination of the point-like contribution from the resonance location and that from the associated wave-propagation zone. In the case of the MR we find, similarly to \citet{Terquem03} for the pure-$B_\varphi$ field, that the point-like contribution to $T_m$ typically has a smaller magnitude than the contribution from the surrounding region, and that these two contributions can have opposite signs. (For the most part, the direction of the point-like torque oscillates at low values of $m$ and converges toward a set sign at high $m$.) The same behavior holds true for the AR, but in general the point-like contribution from the AR is much smaller in magnitude in comparison with both the contribution of the surrounding AR region and the point-like MR torque. 

Table~\ref{tab1} also shows the breakdown of the integrated torque for a field that is twice as strong, i.e. $\beta_{z}=0.4$, at $m=20$. Equations \eqref{MRNobrg} and \eqref{ARNobrg} indicate that, when the field becomes stronger, the magnetic and Alfv\'en resonances move further away from the planet, which has the effect of decreasing the magnitude of the planetary potential at their locations. We find, however, that while the strength of the disk response (as measured by the magnitude of $\Im\{W^\prime\}$) indeed decreases in the AR region (as well as in the FMS wave propagation region beyond the outer Lindblad turning point) when $B_{z0}$ is increased, other factors conspire in this case to increase the enthalpy amplitude in the MR region. We have verified this trend with numerical simulations, which show a more defined wake from SMS waves but a weaker Lindblad wake when $\beta_z$ is decreased. The decrease in torque from the Alfv\'en and FMS wave propagation regions even as the contribution from the MR region goes up results in a slight reduction in the total torque when $\beta_{z}$ is decreased from 0.8 to 0.4. Since the torque at a given value of $m$ is not always representative of the cumulative torque (typically on account of an ancillary effect such as the variation in the sign of the point-like contribution described in the preceding paragraph), we calculated the torque for the $\beta_{z}=0.4$ case over the same range of $m$ and $k_{z}h$ that was employed in Figure~\ref{PBzTorque}. We found that the total torque calculated in this way was again slightly smaller for the stronger field. We also carried out numerical simulations for the two values of $\beta_z$ considered in Table~\ref{tab1} and confirmed that the torque is reduced for the higher value of $B_z$.

The numerical estimates given in the caption to Figure~\ref{PBzTorque} indicate that, for $\beta_z=0.8$, the total torque on the planet is roughly the same in 2D and 3D. If this inference is correct, the implied behavior would contrast with that of HD disks, in which the 2D torque dominates \citep[e.g.,][]{TanakaEtal02}. \citet{MutoEtal08} inferred that the 3D torque exerted on a given side of the planet exceeds the corresponding 2D torque in the limit $\beta_z \ll 1$ (for which the MR contribution dominates), but this regime is not relevant to the formation region of giant planets. Furthermore, their analysis was carried out in the context of the shearing-sheet approximation and therefore could not anticipate our findings (for $\beta_z \lesssim 1$) that the 3D torque decreases only slightly with increasing field strength for given values of $k_zh$ and $m$ because of the opposing effects of the two sides of the planet, and that, in addition, the $k_z\ne 0$ contributions to the total torque
largely cancel each other out.\footnote{We do, however, find evidence for the trend that \citet{MutoEtal08} identified --- that, as $\beta_z$ decreases, the evanescence regions grow and the point-like contribution of the MRs goes up --- also in the higher-$\beta_z$ regime that is of interest to us (see Table~\ref{tab1}).}

\subsection{$\partial B_{\varphi}/ \partial z \ne 0$, $B_z=0$ Field Configuration} 
\label{partialBphi}

We consider this field configuration both here and in Paper~I even though it is not realistic in an attempt to isolate the effect of the $\partial B_\varphi/\partial z$ term on planet migration in wind-driving disks. Figure~\ref{PBphiG} shows the net torque $T_m$ vs. $m$ for this configuration in the 2D limit and for the two vertical wavenumbers considered in Figure~\ref{PBzTorque}. As in the pure-$B_z$ case shown in the latter figure, we were unable to derive results for a few low-$m$ modes in the $k_zh=3.12$ case, but their effect is likely not crucial in this case either. We find that the contributions from the two $k_z\ne0$ integrals that we present roughly cancel each other out as they did in the pure-$B_{z}$ case, and we thus approximate the total torque by the contribution of the $k_{z}=0$ mode. In this way we infer a value for the  2D torque that is nearly equal to that in the HD limit, which agrees with the result of the numerical simulations presented in Paper~I (see Figure~I.10).

As we discussed in Section~\ref{ResTPresults}, MRs appear above and below the midplane in the 3D solutions for this case, and the evidence for them --- seen in the behavior of the perturbed enthalpy --- persists into the 2D limit.  In this limit, the MRs can be regarded as the ``smeared out'' (over the vertical extent of the disk) version of the 2D, pure-$B_\varphi$ resonances found by \citet{Terquem03}. These ``diluted'' MRs are, however, quite weak, and their contribution to the torque is very small. In addition, there are no ARs for this field configuration in the 2D limit. This is because the terms in the coefficients of Equation \eqref{Diff} that would have given rise to these resonances cancel out in the limit $k_z,\, B_z \rightarrow 0$.\footnote{This is not apparent from the results given in Section~\ref{methods}, where, to simplify the presentation, we do not write down the expressions for the coefficients of Equation \eqref{Diff}. However, in the related case of a pure-$B_\varphi$ field considered in Paper~I, we show explicitly that the ARs vanish in the 2D limit (see Equation~I.15).} These facts explain our finding that the magnitude of the 2D torque for this case is close to its HD value.

We also already pointed out (in connection with Equation \eqref{ARNobrg}) that, in 3D, ARs only exist in this case for $m^2>3r^2/h^2$. When this condition is satisfied, the ARs appear very close to the planet, even more so than the MRs (see Figure~\ref{Reslocation}). The larger the value of $h$, the smaller the critical value of $m$, and since for low values of $m$ the planet's potential couples well to the disk (see Equations \eqref{poten} and \eqref{LaplaceC}), the stronger could be the influence of the planet on the disk. In the example presented in Figure~\ref{PBphiG} we chose the disk half-thickness to be of the order of what would have been the density scale height ($c/\Omega$) had we explicitly included the vertical stellar gravity. This was sufficient to guarantee that, despite the fact that we assumed a uniform-density disk, the effect of the ARs on the integrated torque did not become artificially large.

\subsection{ $B_{z}\ne 0$, $\partial B_{\varphi}/ \partial z \ne 0$ Field Configuration} \label{BzBphiResults2} 

Having considered the effects of the $B_z\ne 0$ and $\partial B_{\varphi}/ \partial z \ne 0$ field configurations separately in the previous two subsections, we now investigate the outcome of combining them together. As we discussed in connection with Equation \eqref{Vor}, the simultaneous presence of these two terms in the angular momentum conservation equation gives rise to vertical angular momentum transport in a magnetically threaded disk, and they are thus a key ingredient of a wind-driving disk model. We parametrize this configuration through the values of $\partial B_{\varphi}/ \partial z$ (which equals $B_{\varphi 0}$; see footnote~\ref{note1}) and $B_{z0}$ (or, equivalently, of $\beta_z$ and $\beta_{\varphi}$). The ratio of the azimuthal field amplitude at the disk surface to the vertical field component can be expressed as  $|B_{\varphi \rm s}/ B_{z0}| = h \sqrt{\beta_z/\beta_\varphi}$, and is equal to 0.13 for our chosen parameters. This value is consistent with models of wind-driving disks that are magnetically active at the midplane, for which this ratio is typically inferred to be $\ll 1$ \citep[e.g.,][]{WardleKonigl93}.
 
\begin{figure*}
\begin{center}
\includegraphics[ totalheight=.55\textheight, angle=0]{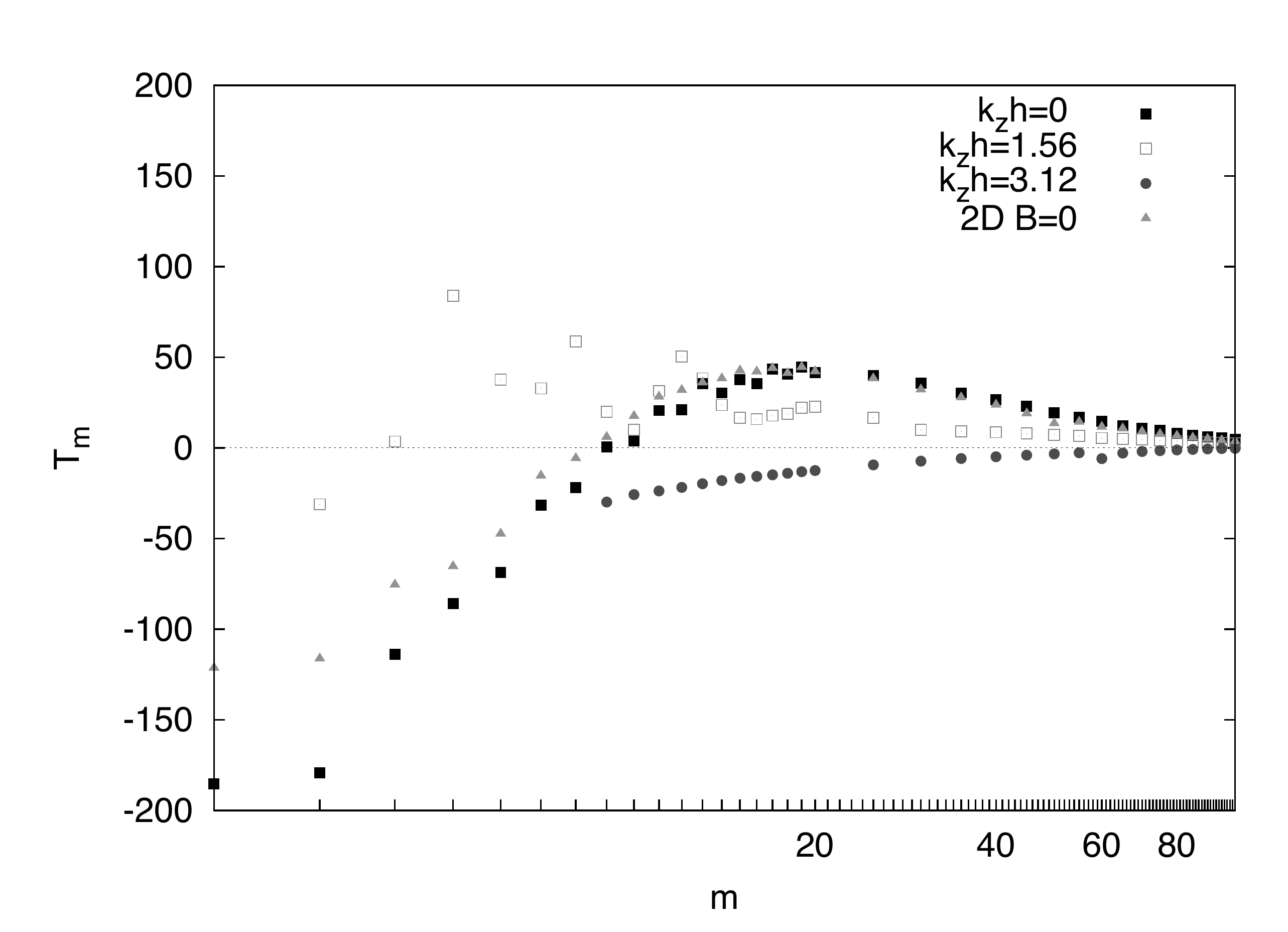}
\caption{Same as Figure~\ref{PBzTorque}, but for a $\partial B_{\varphi}/\partial z \ne 0$, $B_z=0$ field configuration. In this example, $\beta_\varphi=0.04$.
For the $k_{z}h=3.12$ case we were able to obtain numerical results only for $m>8$, and, for $k_{z}h=1.56$, only $m>2$.  Extrapolating where necessary, we find the cumulative dimensionless torque for the $k_{z}h=0$, 1.56, and 3.12 cases to be $\sim$ 1200, 1000, and -900, respectively. Thus neglecting the contribution from higher $k_{z}$ modes, we estimate that the total dimensionless torque is similar to the value for the HD case ($\sim 1300$).}  \label{PBphiG} .
\end{center}
\end{figure*}

\begin{figure*}[htp!] 
\centering 
\subfigure{ \includegraphics[ width=2.in, height=3in, angle=270]{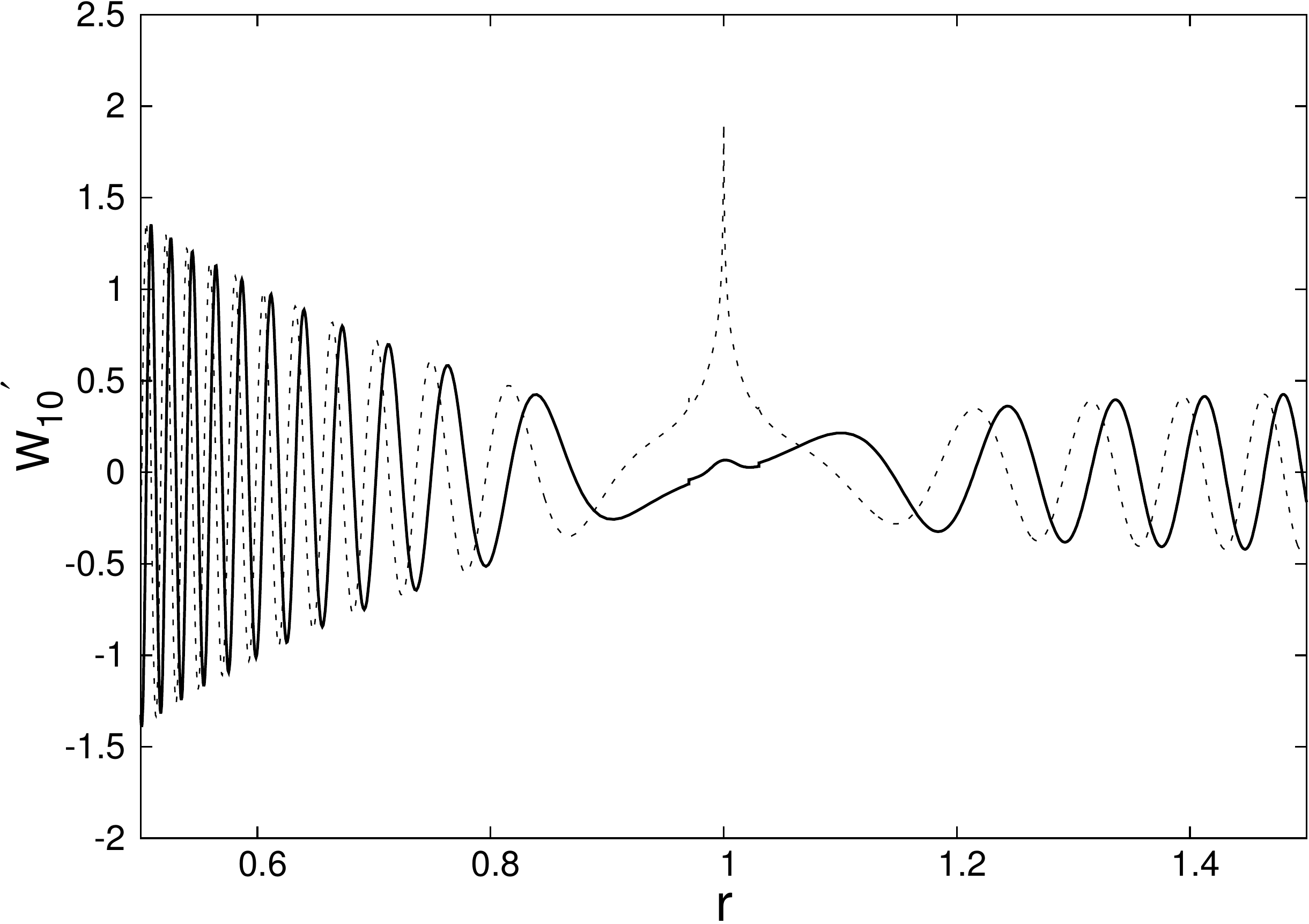}} \quad
\subfigure{ \includegraphics[width=2.in, height=3in, angle=270]{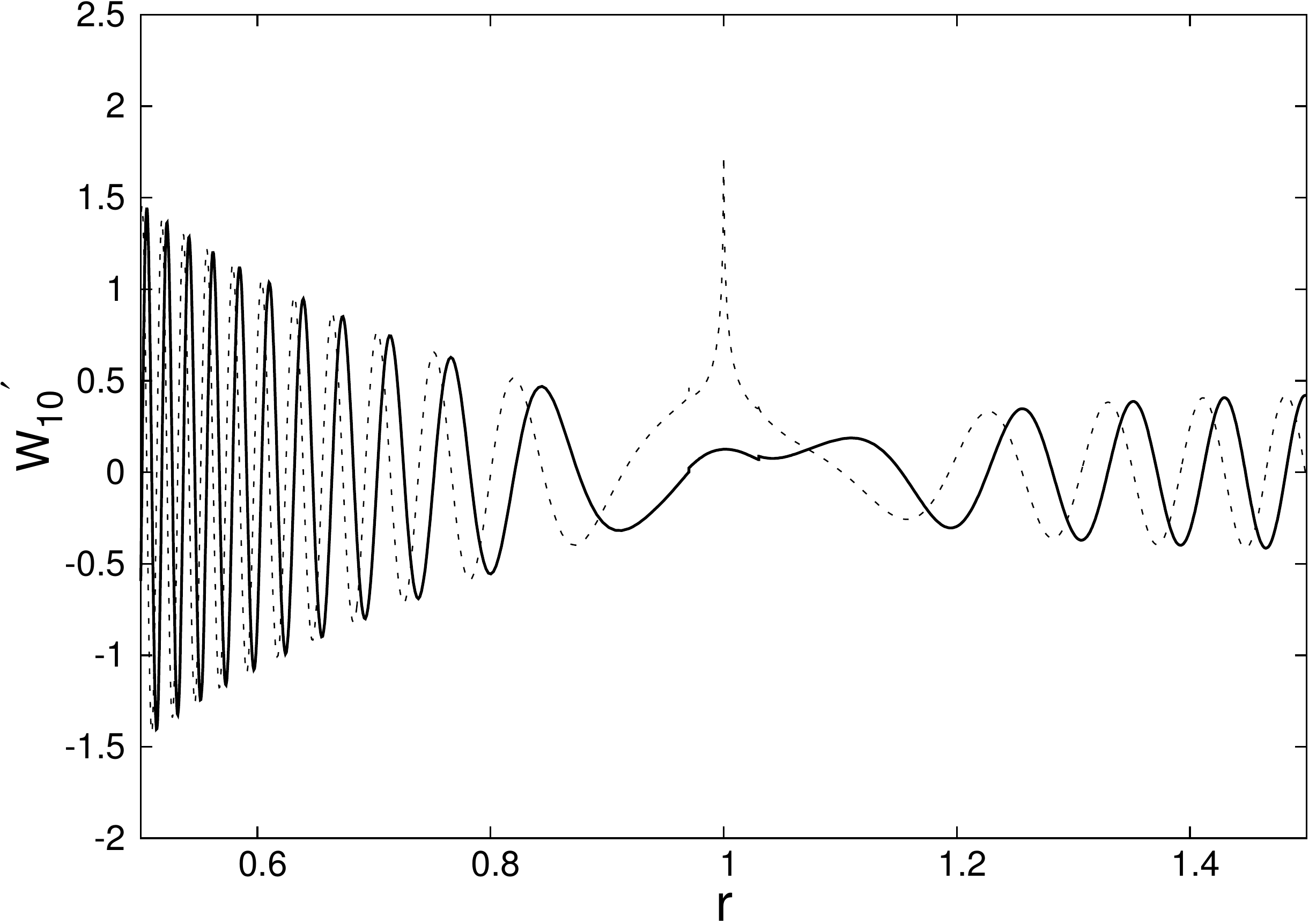} }
\caption{Spatial behavior of the real (dashed line) and imaginary (solid line) parts of $W^\prime_{10}$, the perturbed enthalpy for the $m=10$ mode, in the 2D limit of the $B_{z}\ne 0$,  $\partial B_{\varphi}/\partial z \ne 0$ field configuration. Results are shown for a fixed vertical field amplitude ($\beta_z=0.8$) but two different azimuthal field parameters:
$\beta_\varphi = 0.04$ (left) and $\beta=0.001$ (right, corresponding to a $\sqrt{40}$ times higher value of $|\partial B_{\varphi}/\partial z|$). 
The small sharp peaks exhibited by the dashed curve near $r=1$ correspond to the order-$h^2$ terms in the expression (Equation  \eqref{MRNobrg}) for the MRs and have negligible effect on the torque. The normalized integrated torque $T_{10}$ is $\sim 100$ for the left panel and $\sim 300$ for the right one, indicating that the magnitude of the torque increases with the strength of the field gradient.}
\label{BpGwave} 
\end{figure*}

\begin{figure*}
\begin{center}
\includegraphics[ totalheight=.55\textheight, angle=0]{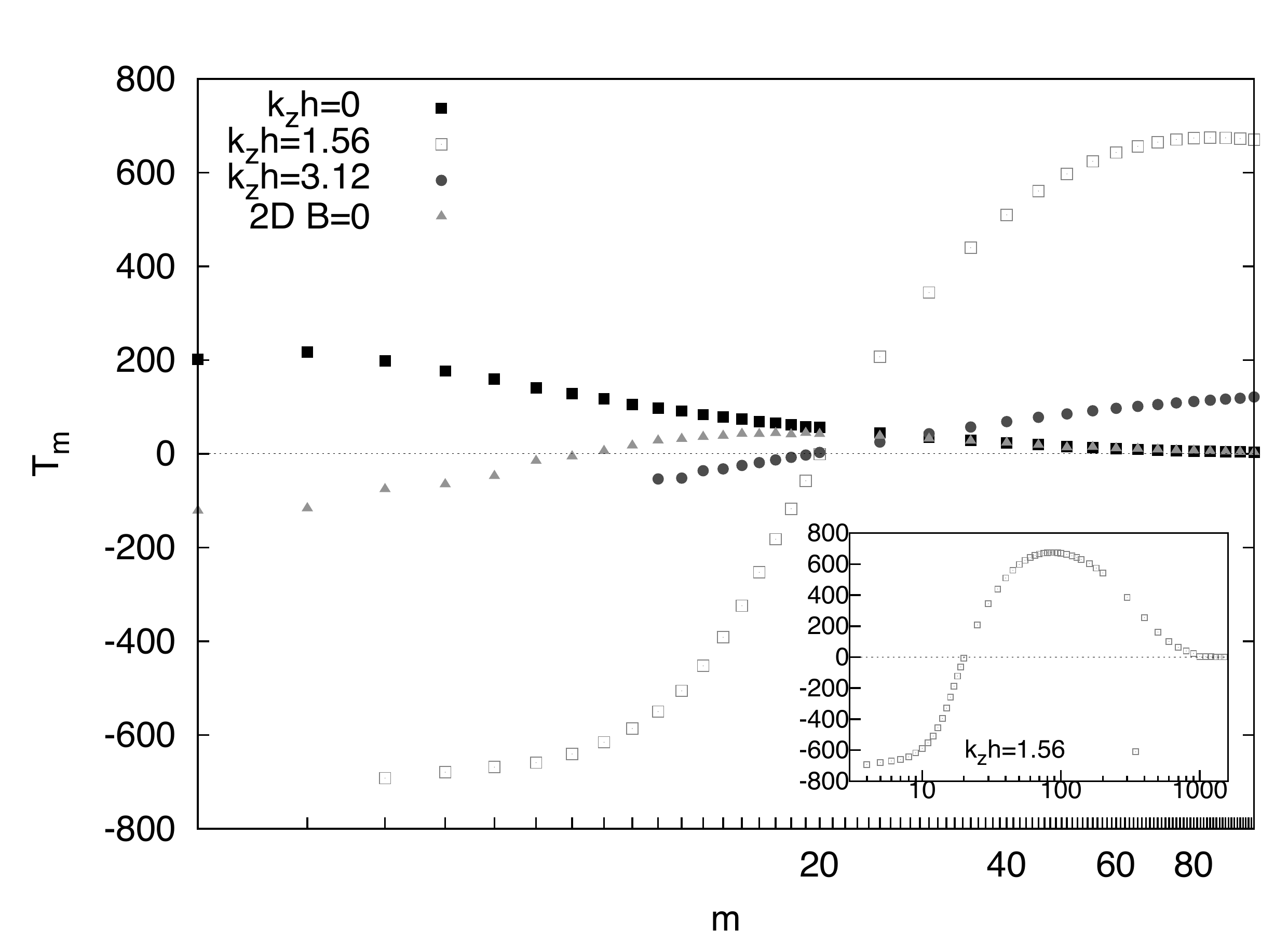}
\caption{Same as Figure~\ref{PBzTorque}, but for a $\partial B_{\varphi}/\partial z \ne 0$, $B_z\ne 0$ field configuration. In this example, 
$\beta_{z}=0.8$ and $\beta_\varphi = 0.04$. We estimate a  torque for the 2D $k_z =0 $ mode of $\sim$ 3400 in dimensionless units,  nearly 3 times larger than the torque in the HD case ($\sim 1300$). The $k_z\ne 0$ modes converge to zero much more slowly with increasing azimuthal mode number $m$. The inset shows the result for the (dominant) $k_z h = 1.56$ mode, which yields an integrated dimensionless torque of $\sim 230,000$, a factor of $\sim 200$ larger than the HD torque.}
\label{BpgTorque}
\end{center}
\end{figure*}

Not surprisingly, the behavior of the ``combined'' field configuration differs from that of its separate components. We noted in Section~\ref{ResTPresults} that one distinctive feature of adding the $B_\varphi$ gradient term to the uniform $B_z$ term in the 2D limit  is the emergence of a 
 wave propagation region in the immediate vicinity of the planet (see top panel of Figure~\ref{WavePropBphiG}). We illustrate the detailed structure of this region in Figure~\ref{BpGwave}, which plots the radial dependence of the perturbed enthalpy $W^\prime$ for two different values of $\beta_\varphi$ (while keeping the magnitude of $\beta_{z}$ as well as the value of $m$ fixed). It is apparent from the shape of the imaginary component of $W^\prime$ (which enters into the torque expression, Equation \eqref{T_mA}) that these solutions exhibit wave-like behavior near the planet, and that this behavior becomes more pronounced as $|B_{\varphi 0}/B_{z0}|$ increases. This region can be expected to make a strong contribution to the torque, and we indeed find (see Figure~\ref{BpgTorque}) that the integrated torque in this case continues to grow with decreasing $m$, and that, correspondingly, the cumulative torque is significantly larger than the value for either of the 2D cases shown in Figures~\ref{PBzTorque} and~\ref{PBphiG}. Another noteworthy finding is that the contributions to the torque from this region have the same (positive) sign on both sides of the planet. This behavior, which was not previously encountered in planet--disk interactions, is also exhibited by the $k_z\ne0$ modes for this field configuration (see next paragraph).

As the bottom panel of Figure~\ref{WavePropBphiG} illustrates, the distribution of the wave propagation regions for the $k_z \ne 0$ modes of this field configuration is qualitatively similar to that of a pure-$B_z$ field. In fact, given the smallness of the factor $h^{2}\partial v_{{\rm A}\varphi }^{2}/v_{{\rm A}z}^2=B_{\varphi \rm s}^2/ B_{z0}^2\approx 0.02$  in Equations \eqref{MRNobrg} and \eqref{ARNobrg}, the locations of the MRs and of the ARs are essentially the same in these two cases. However, the disk response is very different. This is demonstrated by the last two lines in Table~\ref{tab1}, which provide the breakdown of the contributions to the integrated torque in this case for the same two values of $m$ as in the first two lines of this table, which list the corresponding contributions for the pure-$B_z$ field configuration (see Section~\ref{pureBz}). It is seen that, whereas the contribution of the MR regions increases with $m$ and that of the AR regions decreases with $m$ in both cases, other major aspects of the disk behavior have changed. In particular, while the regions interior and exterior to the planet contribute with opposite signs in the pure-$B_z$ case (as they also do in an HD disk), in this case the contributions from the MR and AR regions have the same sign on both sides of the planet for sufficiently high values of $m$ ($\gtrsim 20$ in this case): Those from the MRs are positive everywhere, whereas those from the ARs change from negative to positive as $m$ increases (and we verified that in each case the point-like contribution has the same sign as that from the surroundings). Thus, in contrast with the behavior of the pure-$B_z$ field, where the contributions from the two sides of the planet offset each other and lead to comparatively small integrated torques, in this case the growth of the MR torque with increasing $m$, and the fact that the contributions from the two sides of the planet add up, 
can result in very large integrated torques.

Calculating the cumulative torque for the $k_z\ne 0$ modes in this case is complicated by the fact that the values of $T_m$ are still growing even as $m$ approaches $\sim 100$ (see Figure~\ref{BpgTorque}). Integrating $\partial T_m/\partial A$ at high values of $m$ is computationally challenging because of the high oscillation frequency, which necessitates taking ever smaller step sizes. We have overcome this difficulty by reducing the integration-domain boundaries for high values of $m$, making however sure that, with each such reduction, the torque already calculated for a given value of $m$ did not change. The inset in Figure~\ref{BpgTorque} shows the result of this calculation up to $m=1500$ for the lowest vertical mode ($k_zh=1.56$). Since, for high values of $m$, the contribution from the regions farthest away from the planet (the ``Lindblad'' and AR regions; $T_{Li/o}$ and $T_{ARi/o}$ in Table~\ref{tab1}) are negligible, the value of the integral is not meaningfully affected by shrinking the integration domain. At these very high values of $m$, the contribution from the MRs --- which is added with the same sign from both sides of the planet -- dominates. However, as the mode number continues to increase, the physical width of the resonance becomes progressively smaller (corresponding to a decreasing width of the wave propagation region, akin to the high-$m$ behavior of the pure-$B_z$ field seen in Figure~\ref{WavePropBphiG}), and eventually (for $m \gtrsim 700$) the point-like contribution becomes negligible and $T_m$ converges to zero. The behavior of the other vertical modes is similar and they evidently also contribute a net positive torque (the case $k_zh = 3.12$ is shown in Figure~\ref{BpgTorque} for $m\le100$); however, since the contribution of the $k_zh=1.56$ mode is by far the dominant one, we have not pursued the higher $k_z$ modes to full convergence.

\begin{table*}\centering
\caption{
Net torques for different field configurations in 3D ($k_{z}h=3.12$)}
\ra{1.3}
\begin{tabular}{@{}crrrcrrrcrrrc@{}}\toprule
 $\textbf{B}$ & $\beta_{z}$ &  $\beta_{\varphi}$ &$m$ &&  $T_{Li}$ & $T_{ARi}$ & $T_{MRi}$  & $T_{MRo}$ & $T_{ARo}$ & $T_{Lo}$  & $T_{\rm integrated}$ 
 \\ \midrule
 $ B_{z}  $ &0.8  & ----- & 20 &&           -0.5 & -16.5 & -40.7 & 32.1 & 11.4 & 0.2 & -14.0\\
  $ B_{z} $ &0.8  &  ----- &100&&            -0.1& -2.3 & -65.2 & 63.6 & 2.2& 0.1  & -1.7 \\
  $B_{z} $ &0.4 &  ----- &20&&    $>-0.01$ &  -2.2 &    -54.5&   43.0 & 0.1 &  $<0.01$ & -13.6\\
$B_{z} + \partial B_{\varphi} / \partial z $ & 0.8   & 0.04&  20&&             -0.5 & -20.4 & 15.3 & 15.3& -7.0 & 0.1 & 2.8\\
$B_{z} + \partial B_{\varphi} / \partial z $ & 0.8  &  0.04&100&&           -0.2& 0.2& 60.9& 59.7& 0.8& 0.1 & 121.5\\
\label{tab1}
\end{tabular}
\end{table*}

\section{Discussion} \label{discussion}

The main goal of the present analysis has been to gain insights into Type~I planet migration in wind-driving, magnetized disks. A key property of such disks is vertical angular momentum transport, which is brought about by the magnetic torque term $\propto B_z \partial B_\varphi/\partial z$. Given that the equilibrium structures of such systems cannot be studied self-consistently in the context of ideal MHD, we considered separately the behavior of pure-$B_z$ and pure-$\partial B_\varphi/\partial z$ disks, and then attempted a linear perturbation analysis also for the combined $B_z + \partial B_\varphi/\partial z$ field configuration.

In the case of a uniform vertical field component, we determined that the torque is reduced in comparison with the pure-HD torque
in both 2D and 3D. This implies that planet migration would still be inward, but at a lower (by a factor of 
$\sim 2$) rate than in a disk not threaded by a large-scale field. For the case of a uniform azimuthal field in 2D, \citet{Terquem03} and \citet{FromangEtal05} found that the torque is actually enhanced over the HD torque (resulting in faster inward migration). However, these authors also demonstrated that if $|B_\varphi|$ decreases sufficiently rapidly with radius then inward migration can be strongly suppressed for ($B_\varphi(r) \propto r^{-1}$) and even reversed (for $B_\varphi(r) \propto r^{-2}$). The physical reason for this behavior is that an outward decrease in the field amplitude reduces the contribution of the outer MR region (where the torque exerted by the planet is positive) relative to that of the inner MR region (which has the opposite effect). An outward-decreasing field amplitude (corresponding to $q_z<0$; see Section~\ref{Equilibrium}) can be expected to arise naturally in the case of an open magnetic field that is dragged inward by the disk accretion flow, and it is thus of interest to inquire whether a similar reversal in the direction of planet migration is possible also for the astrophysically more relevant case of a vertical midplane field. We now show that the answer to this question is no.

In the 2D limit, a stronger vertical field leads to a decreased torque (see Section~\ref{pureBz}), in contrast with the pure $B_\varphi$ case, where the converse is true. Thus, if $B_{z}$ decreases with $r$, the torque from the inner region of the disk ($r<1$) will be even weaker relative to the torque from the outer region ($r>1$) than in the uniform-$B_z$ case, resulting in a more positive overall torque and hence in faster inward migration. An added effect arises from the fact that a negative value of $\partial B_{z} / \partial r$ corresponds to an outward-decreasing magnetic pressure that acts to counter the stellar gravity, thereby shifting the corotation radius inward (see Equation~(I.17)). This, in turn, places the outer magnetic resonances and associated turning points closer to the planet, resulting in a further enhancement of inward migration. 
In the 3D case, our numerical simulations and semianalytic results indicate that the total torque 
similarly decreases with increasing field strength for the range of $\beta_{z}$  values we consider, and that the magnitude of the torque is comparable to that in 2D and is again lower than the HD torque. This leads us to infer that it is unlikely that a realistic radial gradient in $B_z$ could give rise to outward migration.

In the case of a pure $ \partial B_{\varphi}/ \partial z$ field configuration, we found that the effect of the magnetic field on the planet--disk interaction is minimal. In fact, the behavior of such a disk is very similar to that of an HD disk, although it also exhibits weak magnetic (in 2D and 3D) and Alfv\'{e}n (in 3D) resonances away from the midplane. However, when both a vertical field and a growing azimuthal field component are present, we discovered that the torque exerted by the planet on the disk is greatly enhanced compared to the HD case. In the 2D limit, our turning-point analysis for this configuration (Section~\ref{ResTPresults}) indicates that wave propagation occurs in the vicinity of the planet. The torque on this close-in region dominates the interaction and leads to inward migration that is faster by a factor of $\sim 3$ than in an HD disk. The $k_z\ne 0$ modes also contribute to inward migration, but the numerical evaluation of their cumulative torques is complicated by the fact that the integrated torque $T_m$ for each of these modes converges to zero only at very high values of $m$ ($\gtrsim 10^3$). We pursued a full calculation only for the lowest (but, by far, the dominant) vertical mode ($k_zh=1.56$), which yielded a cumulative torque that is a factor of $\sim 200$ larger than the HD value. In this case the torque is dominated by the contribution of the MRs, which move closer to the planet as $m$ is increased. The origin of the comparatively strong torque in both the 2D and 3D regimes is the fact that, unlike the situation in the HD and pure-$B_z$ cases, where the torque contributions from the resonance regions on the two sides of the planet add with opposite signs and partially cancel each other out, for this field configuration they combine with the same sign. This, in turn, suggests that a radial gradient in the field strength would only affect the speed of the planet's radial drift in this case, but not its (inward) direction. In Appendix~\ref{appen} we show that the behavior of the disk for this field configuration is tied to the term $\propto B_{z} \partial B_{\varphi}/ \partial z$ in the angular momentum conservation equation. Thus, in this case the planet evidently taps directly into the vertical angular momentum transport channel that enables radial accretion (i.e., $v_{0r}< 0$) in the equilibrium disk model, and this becomes the main mechanism through which the planet loses angular momentum and moves (very rapidly) inward.

\section{Conclusion}\label{conclusion}

This paper and its companion (Paper~I) explore the effects of a large-scale, ordered magnetic field in protoplanetary disks on Type~I planet migration. A large-scale, open field can be expected to be present in the planet formation regions of such disks on account of the advection by the disk accretion flow of the interstellar field lines that thread the parent molecular cloud core. The inward dragging of the field lines would generate a radial field component, and the latter would, in turn, be sheared by the differentially rotating disk material and give rise to an azimuthal field component. This would naturally give rise to a magnetic torque ($\propto B_z \partial B_\varphi/\partial z$) and hence to vertical angular momentum transport, possibly in conjunction with the development of a centrifugally driven disk wind. Our primary motivation in this work is to determine the behavior of planets that reside in disk regions of this type. This task is challenging because quasi-steady equilibrium models of disk regions that are subject to vigorous radial advection and azimuthal shearing require the application of nonideal MHD, which is a complex undertaking. In preparation for a full treatment along these lines, we set out to investigate simpler model problems that can be treated using ideal MHD. In Paper~I we examine a range of problems of this kind through a combination of numerical simulations and a linear perturbation analysis. This paper complements that work by focusing on the semianalytic approach and applying it to the $B_z+\partial B_\varphi/\partial z$ field configuration, which, despite being at the heart of the wind-driving disk model, cannot be simulated using an ideal-MHD numerical code.

While the calculations undertaken in this paper employ a number of simplifications, we present a general formulation of the perturbation analysis that could serve as a reference for a more complete future study. In particular,  we incorporate the equilibrium inflow speed $v_{0r}$ into the linearization scheme. We also discuss the expected effect of including the equilibrium magnetic tension term ($\propto B_z \partial B_r/\partial z$), another key component of the wind-driving disk model, and show that it could lead to additional asymmetric resonances and thereby potentially modify the differential torque. However, our applications concentrate on the magnetic terms that are relevant to vertical angular momentum transport, $B_z$ and $\partial B_\varphi/\partial z$, which we first consider separately and then jointly. Our semianalytic scheme is patterned on the work of \citet{Terquem03}, who focused on the case of a pure-$B_\varphi$ field in 2D. (We consider this case and its generalization to 3D in Paper~I.) For any given field configuration, we identify the resonances and the turning points of the governing second-order differential equation and locate the wave propagation regions where the bulk of the interaction between the planet and the disk takes place. In calculating the torque, we take account of both the point-like contribution from the resonance location and the contribution from the surrounding wave-propagation region.
For a magnetized disk, the relevant waves are SMS and Alfv\'en (associated with the magnetic and Alfv\'en resonances, respectively) as well as FMS. The latter generalize the acoustic waves that propagate away from the effective Lindblad resonances (which are, in fact, turning points) in an HD disk.

The pure-$B_z$ case was previously studied by \citet{MutoEtal08}. However, they employed a shearing-sheet formulation, which does not enable an evaluation of the differential torque. In addition, they modeled the 3D interaction by using the WKB approximation, which  prevented them from identifying the full set of turning points for this case, and by considering the limit $\beta_z \ll 1$, which corresponds to an unrealistically strong field. Thus, even though \citet{MutoEtal08} identified the relevant waves and resonances for this problem, their 3D results are of limited applicability and cannot always be directly compared with those obtained using our more general formulation.
In the 2D limit we confirmed these authors' finding that the torque is reduced in comparison with the HD regime as the field strength increases, which would have the effect of slowing down inward migration. We pointed out that, in contrast to the case of  a pure-$B_\varphi$ field in 2D, for which \citet{Terquem03} and \citet{FromangEtal05} showed that an outward decrease in the field amplitude would tend to reduce the inward drift and might even reverse it, for the pure-$B_z$ case such a decrease would have the opposite effect, acting to speed up inward migration. In 3D we found that the contributions of the $k_z \ne 0$ modes roughly cancel each other out for typical wind-driving disk parameters, so that the total torque is comparable to that in the 2D limit, again implying a reduced, but still inward-directed, migration. 

We found that a stand-alone $\partial B_\varphi/\partial z$ field configuration has little effect on planet migration but that the torque from a disk in which both $B_z$ and $\partial B_\varphi/\partial z$ are nonzero is dramatically increased in comparison with the pure-$B_z$ case. In the 2D limit, this increase is associated in part with the appearance of a wave propagation region much closer to the planet, where the planet's gravitational influence is correspondingly greater. Thus, for a vertical field parameterized by $\beta_z=0.8$, the (dimensionless) total torque increases from $\sim700$ in the pure-$B_z$ case (as compared with $\sim1300$ in an HD disk) to $\sim3400$ when a comparatively weak $B_\varphi$ gradient term (corresponding to a surface field $|B_{\varphi \rm s}| = 0.13\, B_{z0}$) is added. 
In contrast with the pure-$B_z$ case, the net 2D torque appears to increase with the field strength when the gradient term is added. This suggests that inward migration, which speeds up for a uniform field distribution, may be reduced if the field amplitude decreases sufficiently fast with radius, although we concluded that it is unlikely to be reversed given that in this case the contributions to the torque from the inner wave propagation region have the same sign on the two sides of the planet. We did not explicitly pursue this possibility since we found that the two sides of the planet also contribute with the same sign for the magnetic and Alfv\'en resonance regions of the $k_z \ne 0$ modes, where the bulk of the torque is produced, which makes the sign of the total torque insensitive to any spatial variation of the field. For the $k_z\ne 0$ modes we further found that, for high values of the azimuthal mode number $m$, the MR regions become dominant and contribute a strong positive torque (which induces inward migration). By far the largest contribution is produced by the lowest vertical mode ($k_zh=1.56$), for which we obtained a cumulative torque of $\sim 230,000$. We thus infer inward migration at a rate that is $\gtrsim 200$ times faster than in an HD disk. We interpret this result as a manifestation of the ability of the planet to plug into the efficient vertical magnetic-angular-momentum transport mechanism that operates in a disk with this field configuration. This behavior is fundamentally different from that of planets in the standard Type-I migration scenario, which does not involve direct coupling to the disk's underlying angular momentum transport mechanism.

A more precise determination of the torque in a wind-driving disk must await a full, self-consistent treatment within the framework of 3D, nonideal MHD. In particular, a numerical simulation that incorporates the disk's magnetic diffusivity could follow the evolution of a planet in a quasi-equilibrium wind-driving disk and determine not only the implications of vertical angular momentum transport but also those of other relevant effects that were neglected in the current treatment, such as the magnetic tension force and the background accretion velocity. Furthermore, resistive and viscous dissipation effects that become important on small spatial scales could in practice limit the effect of the high-$m$ azimuthal modes  for $k_zh=1.56$ that, according to the ideal-MHD linear analysis, contribute most of the positive torque on the disk. It would also be necessary to determine whether the strong field--matter coupling condition (Elsasser number $\Lambda \gg 1$) that underlies our formulation indeed applies at the midplane of the planet formation region of real protoplanetary disks. In fact, even if the main conclusion  of our study is corroborated by a more detailed investigation, it would be possible for a small planet located in a wind-driving region of such a disk to avoid rapid inward migration if the above condition is not satisfied at that location.

\acknowledgments
 We thank the referee, Takayuki Muto, for very helpful comments and suggestions, and Don Korycansky for valuable input on the numerical integration procedure. This research was supported in part by NASA Headquarters under NASA Earth and Space Science Fellowship Program Grant NNX09AQ89H (A.B.) as well as by NSF grant AST-0908184 and NASA ATP grant NNX13AH56G. The numerical results in this paper were partially carried out at the Midway High Performance Computing Cluster at the Research Computing Center of the University of Chicago. 

\appendix
\section{Wave Propagation when $K$ is Complex}
\label{appen2}
 
When the coefficient $K$ given by Equation \eqref{Kroot} is complex (i.e., if it can be written in the form $K= a + ib$, where $a$ and $b$ are real numbers and $b\ne 0$), the solution always has a wave-like component (i.e., it can be written in the form $C \exp{[(c+ i d)r]}$, where $C$, $c$, and $d$ are real numbers, and $d\ne 0$). Now, $c < 0$ corresponds to wave damping over a characteristic (e-folding) length $\sim 1/|c|$. A rough criterion for wave propagation in the presence of damping is that this length be larger than the size of the wave-formation (resonance) region, which is $\sim 1/|d|$ \citep[e.g.,][]{Artymowicz93}; in other words, a wave propagation region is characterized by $c^2/d^2 \lesssim 1$. 
One can solve for $c$ and $d$ in terms of $a$ and $b$ by plugging the adopted form of the solution into Equation \eqref{lam}. This yields
\begin{equation}  Y \equiv  \frac{c^{2}}{d^{2}} = \frac{ (\sqrt{a^{2}+b^{2}} -a)^{2}}{b^{2}} \ .\label{ratio}\end{equation}

Equation \eqref{ratio} can be used to determine the wave propagation regions for given values of $a$ and $b$ that define the coefficient $K$. One can gain insight into this question by considering the behavior of $Y$ in the limits of small and large $X \equiv b^{2}/a^{2}$. When $X \ll1$, the result depends on the sign of $a$. For $a>0$, $Y \rightarrow \frac {X}{4}$, i.e., $Y \ll 1$, implying wave propagation. This is consistent with the standard hydrodynamical result that corresponds to $\Im \{K\}=0$ (or $X=0$), where wave propagation is inferred to occur when $\Re \{K\}  > 0$. On the other hand, when $X \ll1$ but $a<0$, $Y \rightarrow \frac{4}{X}$ is $\gg 1$, indicating a region of evanescence (again in accord with the behavior of a hydrodynamical disk in the limit $X=0$).

\begin{figure*}
\begin{center}
\includegraphics[ totalheight=.4\textheight, angle=0]{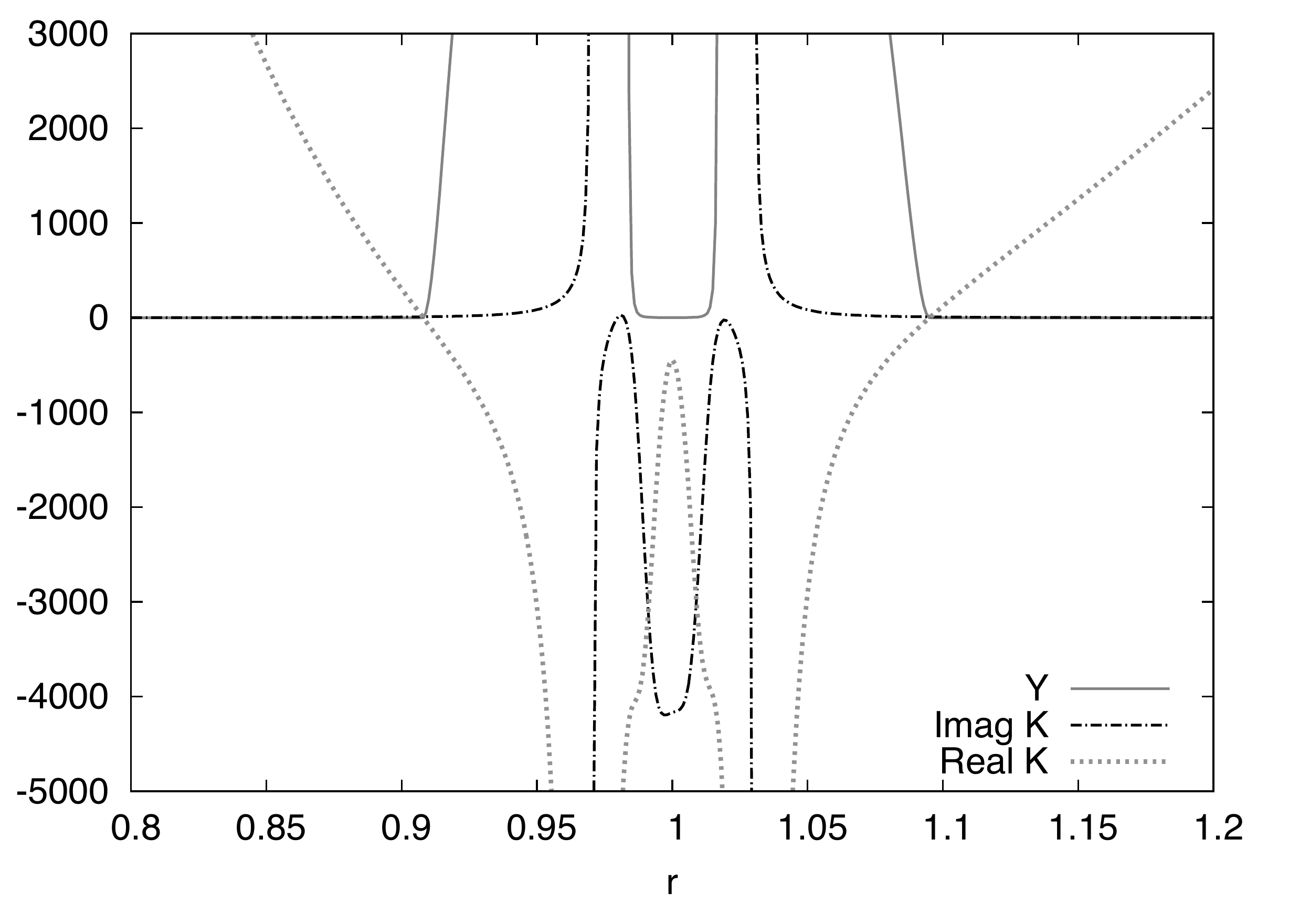}
\caption{Real and imaginary parts of the coefficient $K$ (Equation \eqref{Kroot}) and the ratio $Y\equiv c^{2}/d^{2}$ (Equation \eqref{ratio}) for the 2D limit of the $B_z \ne 0$, $\partial B_\varphi/\partial z \ne 0$ field configuration considered in Section~\ref{BzBphiResults2}. The regions of wave propagation are identified by the conditions
$Y \sim 1$, $|\Im \{K\}| >> |\Re  \{K\}|$
and  $Y \sim 0$, $\Re  \{K\} >0$. The evanescence regions correspond to the locations where  $Y \gg 1$. These results are used in the construction of the schematic shown in the top panel of Figure~\ref{WavePropBphiG}. } \label{cvd}
\end{center}
\end{figure*}

When $X \gg1$, Equation \eqref{ratio} implies $Y \rightarrow 1$, remaining $< 1$ for $a>0$ and $>1$ for $a<0$. In this case the wave propagation criterion formulated above is only marginally satisfied. This case is realized in the vicinity of the planet in the 2D limit ($k_z = 0$) of the $B_z \ne 0$, $\partial B_\varphi/\partial z \ne 0$ field configuration considered in Section~\ref{BzBphiResults2}, as shown in Figure~\ref{cvd}. The profile of the perturbed enthalpy for this model indeed hints at wave-like behavior in that region (see Figure~\ref{BpGwave}), and we have therefore labeled it as a wave-propagation zone in the schematic presented in the top panel Figure~\ref{WavePropBphiG}. We have delimited the extents of the wave-propagation and evanescence regions in this schematic based on the results shown in Figure~\ref{cvd}.

\section{Dominant Torque Terms for the ``Combined'' Field Configuration}
\label{appen}

We wish to determine whether a planet located in a disk with a combined $B_z + \partial B_\varphi/\partial z$ field configuration indeed taps into the vertical magnetic transport of angular momentum (represented by the $\propto B_{z} \partial B_{\varphi}/ \partial z $ term in the angular momentum conservation equation). To this end, we examine the dominant terms in the expression for the perturbed magnetic  torque density acting on the disk, $(r-r_{\rm p})F^{\prime}_{\varphi}$.
From the $\varphi$ component of Equation \eqref{F'CF} we have
\begin{equation}  \mu F^{\prime}_{\varphi}= B^{\prime}_{z} \partial B_{0\varphi}/ \partial z  + i (k_{z}B^{\prime}_{\varphi} - k_{\varphi} B^{\prime}_{z} ) B_{0z}  \label {A1} \, , \end{equation} 
where $k_\varphi = m/r$. In view of our finding in Section~\ref{BzBphiResults2} that the behavior of the disk is different for the $k_{z}=0$ and $k_{z} \ne 0$ vertical modes, we consider these two cases separately. 

\begin{figure*}
\begin{center}
\includegraphics[ totalheight=.4\textheight, angle=0]{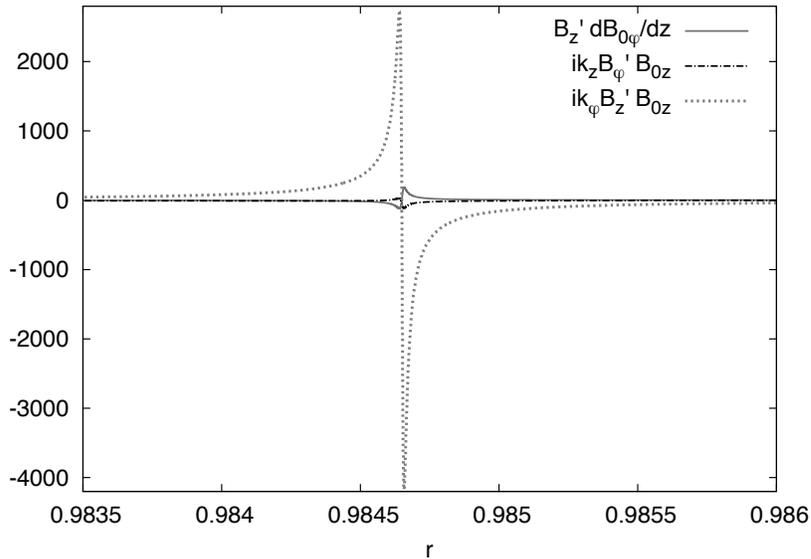}
\caption{Behavior of the (real) contributions to the azimuthal component of the perturbed Lorentz force density (Equation \eqref{A1}) in the vicinity of the inner magnetic resonance for the model parameters that correspond to the bottom entry of Table~\ref{tab1}. The term $\propto k_{\varphi} B_{z}^\prime B_{0z}$ is seen to dominate at and around the resonance.}  \label{BpgVzTorqueTerms}
\end{center}
\end{figure*}

For $k_{z} \ne 0$, the bottom entry of Table~\ref{tab1} highlights the fact that high azimuthal mode numbers $m$ dominate the torque.
It is also clear from this entry that the region around the magnetic resonance provides the main contribution to the torque. Therefore, we consider Equation \eqref{A1} in the MR region and in the limit of a large $k_{\varphi} $. Figure~\ref{BpgVzTorqueTerms} shows the behavior of the three terms on the right-hand side of Equation \eqref{A1} in the inner MR region for the parameters that correspond to the aforementioned table entry: it is seen that the last term, which contains an explicit dependence on $k_\varphi$, clearly dominates.

The $z$ component of the perturbed induction equation (Equation \eqref{ind}) gives 
\begin{equation} i k_{\varphi} r \sigma B^{\prime}_{z} =  \bigg( -\frac{v_{r}^{\prime'}}{r} - \frac{\partial v_{r}^{\prime}}{\partial r} \bigg)B_{0z} - i k_{\varphi}  v_{\varphi}^{\prime} B_{0z}\, . \label{bzp1} \end{equation} 
Substituting the perturbed continuity equation,
\begin{equation}  i k_{\varphi} r \sigma \frac {\rho^{\prime}}{\rho_0} +   \bigg( \frac{v_{r}^{\prime}}{r} + \frac{\partial v_{r}^{\prime}}{\partial r} \bigg)  + i k_{\varphi}v_{\varphi}^{\prime} + i k_{z} v_{z}^{\prime} =0\,  , \label{pcont} \end{equation}
into Equation \eqref{bzp1}, we arrive at the following form for the perturbed vertical field component:
\begin{equation}  B^{\prime}_{z} = B_{0z}\frac {\rho^{'}}{\rho_0} + \frac{ k_{z}}{k_{\varphi} r \sigma} v_{z}^{\prime}B_{0z}\, . \label{bzp2} \end{equation} 
Since we are focusing on the regions around the magnetic resonances (which lie close to the planet), $\sigma$ is generally small, so even for large values of $k_{\varphi}$ one can expect the second term on the right-hand side of Equation \eqref{bzp2} to be the principal component of $B_{z}^{\prime}$. We have verified numerically that this term is indeed the largest factor in the expression for $B_{z}^{\prime}$ (although the contribution of the first term remains nonnegligible). In our analytic formulation, $\rho^{'}/\rho_0 \propto W^{\prime}$ is given by a complicated expression in which
several terms depend on the product $B_{0z} \partial B_{0\varphi}/ \partial z $. However, the dependence of the dominant (second) term in Equation \eqref{bzp2} on this product is more clear-cut. The perturbed vertical velocity, $v_{z}^{\prime}$, is induced primarily by the vertical component of the Lorentz force, which for this field configuration is proportional to $ (B_{\varphi}^{\prime}  \partial B_{0\varphi}/ \partial z) / \rho_0 $. Thus, the leading term that affects the angular momentum transfer from the planet to the disk in this case is (substituting the above dependencies into Equation \eqref{A1}):
\begin{equation}   F^{'}_{\varphi} \propto \frac{k_{z} B^{\prime}_{\varphi} B_{0z}}{\rho_0} \bigg( B_{0z}\frac{\partial B_{0\varphi}}{\partial z} \bigg)\ . \label{Fphi3D}\end{equation} 
It is evident from Equation \eqref{Fphi3D} that the vertical transport of angular momentum by the field plays a major role in shaping the 3D behavior of a disk with this field configuration.

When $k_{z}=0$, the middle term of Equation \eqref{A1} vanishes, and we are again left with terms that depend on the perturbed vertical field (which in this case is given by $B_{z}^{\prime}=  B_{0z}(\rho^\prime/\rho_0)$). Since the interesting behavior of the 2D case illustrated in Figure~\ref{BpgTorque} is found to be at low $m$ (where the inner wave propagation region is largest), we look at Equation \eqref{A1} at small $k_{\varphi}$. In this limit the $ B^{\prime}_{z} \partial B_{0\varphi}/ \partial z$ term is dominant. Accordingly,  $F_{\varphi}^\prime \propto (\rho^\prime/\rho_0) B_{0z} \partial B_{0\varphi}/ \partial z$, demonstrating that the vertical magnetic transport of angular momentum plays a role in the 2D limit as well. 

Our finding that the dominant contributions to the torque for both the $k_z=0$ and $k_z\ne0$ modes have the same sign on the two sides of the planet is consistent with the picture that in the ``combined'' field case the planet loses angular momentum primarily through the large-scale magnetic field. The planet couples to the field indirectly, through the density perturbations that are induced in the gas by its gravitational potential. The perturbed field exerts a back torque that is transmitted to the planet (again, through its gravitational interaction with the gas) even as the field transports the planet's angular momentum to the disk surfaces, where it can be deposited in a centrifugally driven wind (or, alternatively, in torsional Alfv\'en waves that propagate into the ambient interstellar medium). In our formal treatment, the angular momentum transport is described through the launching of MHD waves that propagate both radially and vertically into the surrounding gas. However, the dominant physical transport is effected by the large-scale field that carries the angular momentum in the vertical direction. And while the analysis of the wave propagation regions in the disk is useful for identifying the locations where the planet interacts most strongly with its surroundings, the standard picture in which the torque exerted on the planet depends on the planet's angular velocity relative to the gas (and therefore changes sign between the interior and exterior disk regions) does not apply in this case. In the standard picture, Type~I migration does not depend on the underlying (``viscous'') angular momentum transport mechanism in the disk. The effective disk kinematic viscosity $\nu$ is commonly represented in the form $\nu = \alpha\,c^2/\Omega_{\rm K}$ \citep{ShakuraSunyaev73}, with the parameter $\alpha$ typically taken to be $\ll 1$, and its relative unimportance in the Type-I migration process is due to the fact that the viscous transport term is a factor $\sim \alpha$ smaller than the other terms in the perturbed angular momentum conservation equation. By contrast, for disks in which a large-scale magnetic field dominates the angular momentum transport, the effective value of $\alpha$ is $\gg 1$, so in this case the disk's angular momentum transport mechanism can directly affect the planet's migration.

\bibliographystyle{apj}
\bibliography{mybib2}
\clearpage
\end{document}